\documentclass{ametsocV6.1}

\usepackage{amsmath,amsfonts,amssymb,bm}
\usepackage{mathptmx}%{times}
\usepackage{newtxtext}
\usepackage{newtxmath}
\usepackage{rotating}
\usepackage[finalnew]{trackchanges}

\title{Perturbing parameters to understand cloud contributions to climate change}

    \authors{Margaret L. Duffy\correspondingauthor{Margaret L. Duffy, mlduffy@ucar.edu}, Brian Medeiros, Andrew Gettelman\thanks{Current affiliation: Pacific Northwest National Laboratory, Richland, WA}, Trude Eidhammer}

     \affiliation{National Center for Atmospheric Research, Boulder, Colorado}

\abstract{The sensitivity of cloud feedbacks to atmospheric model parameters is evaluated using a CAM6 perturbed parameter ensemble (PPE). The CAM6 PPE perturbs 45 parameters across 262 simulations, 206 of which are used here. \remove{Notably, the CAM6 PPE is run with a more recent version of CAM6 (CAM6.3) than was used for AMIP (CAM6.0), and has a smaller total cloud feedback (0.56 W m$^{-2}$ K$^{-1}$) as compared to CAM6.0 (0.81 W m$^{-2}$ K$^{-1}$) owing primarily to reductions in low clouds over the tropics and middle latitudes. }The spread in total cloud feedback and its six components across the CAM6 PPE are comparable to the spread across the CMIP6 and AMIP ensembles\add{, indicating that parametric uncertainty mirrors structural uncertainty}. However, the high-cloud altitude feedback is generally larger in the CAM6 PPE than WCRP assessment, CMIP6, and AMIP values. We \remove{further} evaluate the influence of each of the 45 parameters on the total cloud feedback and each of the six cloud feedback components.\note{order of last and next sentences changed} We \add{also }explore whether the\add{ CAM6} PPE can be used to constrain the total cloud feedback, with inconclusive results. \change{Finally}{Further}, we find that despite the large parametric sensitivity of cloud feedbacks in CAM6, \change{the}{a substantial} increase in cloud feedbacks from CAM5 to CAM6 is not a result of changes in parameter values.\add{ Notably, the CAM6 PPE is run with a more recent version of CAM6 (CAM6.3) than was used for AMIP (CAM6.0), and has a smaller total cloud feedback (0.56 W m$^{-2}$ K$^{-1}$) as compared to CAM6.0 (0.81 W m$^{-2}$ K$^{-1}$) owing primarily to reductions in low clouds over the tropics and middle latitudes.} The work highlights the large sensitivity of cloud feedbacks to both parameter values and structural details \change{of}{in} CAM6.}

\begin{document}

%% Necessary!
\maketitle

%%%%%%%%%%%%%%%%%%%%%%%%%%%%%%%%%%%%%%%%%%%%%%%%%%%%%%%%%%%%%%%%%%%%%
% SIGNIFICANCE STATEMENT/CAPSULE SUMMARY
%%%%%%%%%%%%%%%%%%%%%%%%%%%%%%%%%%%%%%%%%%%%%%%%%%%%%%%%%%%%%%%%%%%%%
%
% If you are including an optional significance statement for a journal article or a required capsule summary for BAMS 
% (see www.ametsoc.org/ams/index.cfm/publications/authors/journal-and-bams-authors/formatting-and-manuscript-components for details), 
% please apply the necessary command as shown below:
%
% \statement
% Significance statement here.
%
% \capsule
% Capsule summary here.

%%%%%%%%%%%%%%%%%%%%%%%%%%%%%%%%%%%%%%%%%%%%%%%%%%%%%%%%%%%%%%%%%%%%%
% MAIN BODY OF PAPER
%%%%%%%%%%%%%%%%%%%%%%%%%%%%%%%%%%%%%%%%%%%%%%%%%%%%%%%%%%%%%%%%%%%%%
%

\section{Introduction}

Constraining equilibrium climate sensitivity (ECS) is a crucial step to quantifying warming from anthropogenic emissions. However, doing so has proven difficult, with little change in the estimated ECS range from 1979 until present \citep{ad_hoc_study_group_on_carbon_dioxide_and_climate_carbon_1979, sherwood_assessment_2020, forster_earths_2021}. Global climate models (GCMs) are useful for understanding the climate's response to forcing. However, the spread in ECS values across models remains large, and has, in fact, increased between CMIP5 and CMIP6. From CMIP5 to CMIP6 the upper ECS estimate increased from 4.7K to 5.6K and the multimodel mean ECS increased from 3.3K to 3.9K. Further, 10 CMIP6 models have an ECS greater than that of the largest CMIP5 model (4.7K) \citep{zelinka_causes_2020}.

ECS depends on both the radiative forcing and the radiative feedback. The change in spread in ECS from CMIP5 to CMIP6 is the result of different combinations of radiative forcing and feedback \citep{zelinka_causes_2020}. The radiative feedback includes contributions from both clear-sky and cloud feedbacks. The spread across models and observations is dominated by the spread in cloud feedbacks \citep{sherwood_assessment_2020, zelinka_causes_2020}. Therefore, constraining cloud feedbacks is key to constraining ECS.

The relationship between ECS, forcing, and feedback can be conceptualized using the top-of-atmosphere (TOA) energy budget, which is given by
\begin{equation} \label{eq:F}
    \Delta N = \Delta F + \lambda \Delta T,
\end{equation}
where $\Delta N$ is the net top-of-atmosphere radiative flux anomaly (positive down), $\Delta F$ is radiative forcing, $\lambda$ is the total radiative feedback parameter, and $\Delta T$ is the global-mean near-surface air temperature response. We define ECS by assuming equilibrium ($\Delta N=0$) and a radiative forcing which corresponds to a doubling of CO$_2$,  given by
\begin{equation} \label{eq:dT}
    ECS \equiv \Delta T_{2x\text{CO}_2} = -\frac{\Delta F_{2x\text{CO}_2}}{\lambda}.
\end{equation}
\add{From Equation }\ref{eq:dT}\add{, it is clear that there is a close, inverse relationship between ECS and radiative feedbacks.} \add{In practice, calculating ECS requires running a 2x or 4xCO$_2$ GCM simulation to equilibrium.}\remove{ However, ECS is often diagnosed using $\Delta T_{4x\text{CO}_2}$ divided by 2 where $\Delta F$ corresponds to a quadrupling of CO$_2$ or using the Gregory regression method of .}\remove{From Equation 
 2, it is clear that there is a close, inverse relationship between ECS and radiative feedbacks.} \add{However, o}\remove{O}ther definitions of climate sensitivity also exist, such as effective climate sensitivity and transient climate sensitivity, and tend to be correlated with ECS across GCMs \citep{gregory_new_2004, rugenstein_three_2021}. The radiative feedback $\lambda$ can be further decomposed into cloud radiative feedbacks and the noncloud radiative feedbacks, given by
\begin{equation}
    \lambda = \lambda_{cloud}+\lambda_{noncloud}.
\end{equation}
The noncloud feedbacks include the Planck, water vapor, lapse rate, and surface albedo feedbacks.\remove{ The negative Planck feedback has the largest magnitude of any feedback. As a result, the total feedbacks are overall negative (stabilizing).} \change{The noncloud feedbacks tend to have lower uncertainty than the cloud feedbacks. Here we focus on cloud feedbacks because of their comparatively large uncertainty.}{We focus on cloud feedbacks because their uncertainty is larger than noncloud feedbacks }\citep{sherwood_assessment_2020}.

The total cloud feedback is composed of various cloud changes with warming. Here we separate cloud feedbacks into both their shortwave (SW) and longwave (LW) contributions and into six cloud feedback components as in \citet{sherwood_assessment_2020} (also called the World Climate Research Programme [WCRP] assessment): High-cloud altitude, tropical marine low-cloud, tropical anvil cloud area, land cloud amount, middle latitude marine low-cloud amount, and high latitude low-cloud optical depth. These six cloud feedback components are the result of known processes, and \citet{sherwood_assessment_2020} assess their values and spreads using multiple lines of evidence. The high-cloud altitude feedback is a robustly positive feedback across lines of evidence \citep{sherwood_assessment_2020} and across GCMs \citep{zelinka_evaluating_2022}. The high-cloud altitude feedback results from cloud-top heights rising nearly isothermally which means that LW emission from cloud tops changes less than the increased surface emission, resulting in greater LW flux divergence and enhanced warming (i.e., a positive feedback) \citep{hartmann_important_2002, zelinka_why_2010}. The tropical marine low-cloud feedback is estimated to be a positive SW feedback resulting from reduced low-cloud amounts in the tropics with warming \citep{bony_marine_2005, klein_low-cloud_2017}. The tropical anvil cloud area feedback comes from a reduction in the area of tropical anvil clouds with warming. This reduced area has compensating LW and SW feedbacks, but observational evidence suggests the LW effect is greater than the SW effect for a net negative feedback \citep{hartmann_tropical_2001, williams_observational_2017}. The land cloud feedback is expected to produce a positive feedback of smaller magnitude than any of the three feedback contributions already described. The land cloud feedback is the result of reduced cloudiness due to reduced relative humidity over land \citep{bretherton_cloud_2014}. Low-cloud reductions dominate the radiative effect of the land cloud feedback, resulting in a positive SW feedback \citep{kamae_robust_2016}. Middle latitude low-cloud feedbacks are expected to provide a positive SW feedback due to reductions in low clouds \citep{kay_processes_2014, zhai_longterm_2015, mccoy_change_2017}. The high latitude low-cloud optical depth feedback results from competing influences on high latitude low-cloud optical depths with warming, and therefore has an estimated feedback of 0 W m$^{-2}$ K$^{-1}$ but with substantial spread \citep{tan_observational_2016, ceppi_observational_2016}. These six components are separate from one another and sum to the total feedback if the unassessed component is small, which it is in most CMIP5 and CMIP6 models \citep{zelinka_evaluating_2022}.

Focusing on GCMs, the spread in total cloud feedbacks has increased from CMIP5 to CMIP6, with an increase in spread in the middle latitude low-cloud amount (and scattering) feedback from CMIP5 to CMIP6 \citep{zelinka_causes_2020}. The CMIP5 and CMIP6 multimodel mean cloud feedbacks (total cloud feedback and its six components) all fall within the WCRP assessed ranges, but there is substantial inter-model spread across the total cloud feedback and most of the six cloud feedback components. Further, the tropical marine low-cloud feedback tends to be smaller (a smaller positive feedback) and the tropical marine anvil cloud area feedback tends to be larger (a smaller negative feedback) than the WCRP assessed range. CMIP5 and CMIP6 models whose assessed feedbacks compared most favorably to the WCRP assessed values had total cloud feedback estimates of 0.4 to 0.6 W m$^{-2}$ K$^{-1}$ and ECS estimates of 3 to 4 K \citep{zelinka_evaluating_2022}.

Despite the tremendous impact of clouds on Earth's climate, their subgrid size requires cloud processes to be parameterized in GCMs. \textit{Here we evaluate the sensitivity of cloud feedbacks to atmospheric model parameter values in the Community Atmosphere Model version 6, CAM6, \citep{danabasoglu_community_2020} using a perturbed parameter ensemble (PPE)}. An advantage of using a PPE is that we are able to use parameter sensitivity to identify processes setting the spread in cloud feedbacks.\remove{ The CAM6 PPE is created using CAM6, which is the atmospheric model in CESM2, one of the high ECS models in CMIP6 . However, the CAM6 PPE has a smaller cloud feedback than the CESM2 AMIP and CMIP simulations.} We use the CAM6 PPE and CMIP6 models to address the following questions:
\begin{enumerate}
    \item How does the spread in cloud feedbacks across the CAM6 PPE compare to the spread in cloud feedbacks across CMIP6 models?
    \remove{Why is the total cloud feedback smaller in the CAM6 PPE than in CESM2?}
    \item What is the sensitivity of cloud feedbacks to parameter values in CAM6?\note{changed the order of 2 and 3}
    \item Can the total cloud feedback be constrained using the CAM6 PPE?
    \remove{Are changes in parameters between CAM5 and CAM6 responsible for the increase in total cloud feedback and ECS from CESM1 to CESM2?}
\end{enumerate}

\add{Additionally, there have been substantial fluctuations in the cloud feedback and corresponding ECS values across generations the Community Atmosphere Model (CAM). The cloud feedback and ECS values increased from CAM5 to CAM6 }\citep{gettelman_high_2019}{. However, we find here that the CAM6 PPE default simulation\add{ (CAM6.3)} has lower \add{cloud feedback and ECS }values than previous CAM6 simulations\add{ (CAM6.0)}. Therefore, we also investigate the following questions:}
\begin{enumerate}
\setcounter{enumi}{3}
    \item Are changes in parameter values (tuning) responsible for the increase in cloud feedback from CAM5 to CAM6?
    \item Why does the CAM6 PPE default simulation have a lower cloud feedback than previous CAM6 \change{simulation}{configuration}?
\end{enumerate}

The CMIP, AMIP, and PPE data are described in Section \ref{sec:data}. A description of the cloud radiative feedback calculation is also provided in Section \ref{sec:data}. Questions 1 through 5 are addressed sequentially in Sections \ref{sec:spread} through \ref{sec:low_cf}.\add{ For interested readers, we discuss the relationship between the CAM6 PPE used here and a version which is tuned based on paleoclimate evidence in Section }\ref{sec:paleo}. We discuss and conclude in Section \ref{sec:dc}.

\section{Data description} \label{sec:data}

\subsection{PPE}

The CAM6 PPE is available for use \citep{eidhammer_cesm22-cam6_2022}, is described in detail in Eidhammer et al. 2023 (unpublished manuscript), and is briefly summarized here. The CAM6 PPE consists of 262 CAM6 simulations\footnote{263 simulations were run, but one was discarded due to numerical instability.}. The simulations are atmosphere-only simulations run with CAM6.3 with imposed climatological SST and sea-ice cover (using the F2000climo component set of the model). The prescribed SST and sea-ice fields have an annual cycle but no interannual variability. One simulation is considered a `default' simulation because it uses the parameter values typical of CAM6. The remaining 261 simulations differ in 45 atmospheric parameter values across parameterizations for cloud and aerosol effects. The parameter ranges are chosen by expert judgment to span physical process limits\add{. The chosen parameter ranges are intended to cover the plausible range of realistic values based on expert assessment, and thus span the range of parametric uncertainty}. Parameter values are selected using latin hypercube sampling, meaning that the parameters are varied simultaneously, but randomly and independently from one another \citep{mckay_comparison_1979}. Latin hypercube sampling ensures that for each parameter, the range of possible values is evenly sampled. Here we show results from variations in 43 parameters.\footnote{Not 45 because two pairs of parameter values are perturbed simultaneously: CLUBB C6rt/C6thl and CLUBB C6rtb/C6thlb.} For each of the 262 PPE simulations, we use two experiments: a control or present day simulation, denoted `PD', and a warm simulation with a uniform 4K SST warming, denoted `SST4K'. All PPE simulations are run with the CFMIP observation simulator package (COSP) turned on, which produces joint histograms of cloud-top pressure and cloud optical depth in each simulation \citep{bodas-salcedo_cosp_2011}. \add{All PPE simulations are run for three years each. }Monthly-mean model output is used throughout.

\begin{figure}
    \centering
    \includegraphics[width = \textwidth]{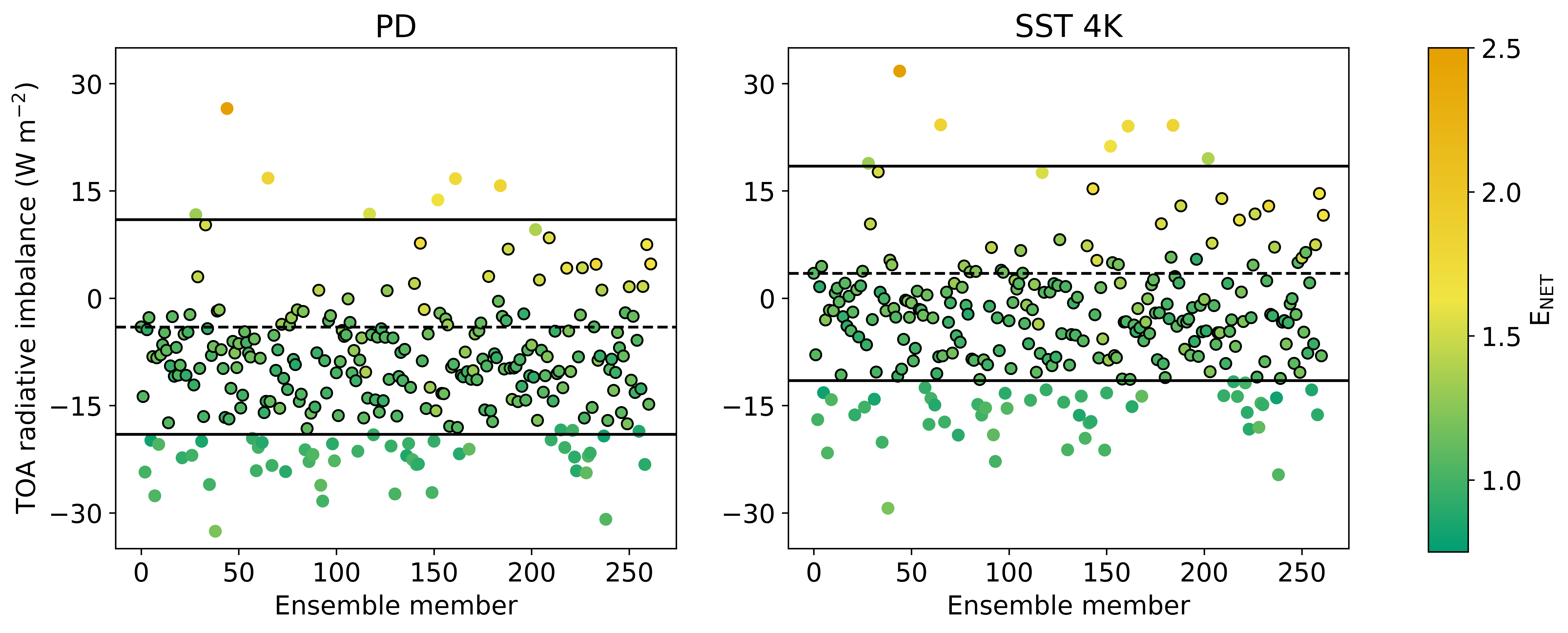}
    \caption{Top-of-atmosphere radiative imbalance in each of the 262 PPE simulations for both the PD (left) and SST4K (right) experiments. The dashed horizontal line goes through the default value and the black lines are 15 W m$^{-2}$ away from the default. Dot colors correspond to the E$_\textrm{NET}$ value in the PD simulation. The simulations in the subset used in the forthcoming analysis are encircled in black.}
    \label{fig:subset}
\end{figure}

The spread in feedbacks across the CAM6 PPE is sensitive to the range of each parameter value chosen in designing the CAM6 PPE. Although the parameter values are chosen by expert judgment to within physical process limits, combinations of different parameter values need not yield realistic climates \citep{stainforth_uncertainty_2005}. To address this, we take a subset of the 206 PPE simulations based on TOA a radiative balance constraint.\add{ None of the CAM6 PPE simulations have been tuned based on TOA radiative imbalance.} The default PD simulation has a TOA radiative imbalance of -4.0 W m$^{-2}$. The default SST4K simulation has a TOA radiative imbalance of 3.5 W m$^{-2}$. Here and throughout the paper, we analyze only the 206 simulations whose TOA radiative imbalance falls within 15 W m$^{-2}$ of the default simulation for both the PD and SST4K experiments. That is, we only keep simulations whose TOA imbalance is between -19 and 11 W m$^{-2}$ in the PD experiment \textit{and} whose TOA imbalance is between -11.5 and 18.5 W m$^{-2}$ in the SST4K experiment. The TOA radiative imbalance for each simulation and experiment is shown in Figure \ref{fig:subset}. \add{The 15 W m$^{-2}$ threshold is chosen based on the data in Figure }\ref{fig:subset}\add{ somewhat subjectively, with the intention of eliminating data in the tails of the TOA imbalance distribution}. We also evaluate our subset using mean-state cloud error for each simulation's PD experiment. \change{W}{To do so, w}e use an aggregated scalar error metric of the impact of cloud errors on TOA radiation in the base state, E$_\textrm{NET}$ \citep{klein_are_2013}, with details as in \citet{zelinka_evaluating_2022}. Reassuringly, we find that our subset eliminates the simulations with largest mean-state cloud errors (Figure \ref{fig:subset}).

\subsection{Feedback calculations}

In the CAM6 PPE, the total radiative feedback is calculated as the ratio of net downward TOA flux $\Delta N $ to surface warming $\Delta T$. We calculate cloud radiative feedbacks using the cloud radiative kernel technique with output from the PD and SST4K simulations. In general, cloud radiative kernels calculate the sensitivity of TOA radiation to perturbations in cloud fraction. Here we calculate cloud radiative feedbacks using CAM5 joint-histogram kernels from \citet{zelinka_computing_I_2012}\footnote{The radiative transfer code is the same in CAM5 and CAM6, so these kernels are accurate for the CAM6 simulations used here.}. The kernels of \citet{zelinka_computing_I_2012} are constructed from joint histograms of cloud-top pressure and optical depth to separate cloud-radiative effects into contributions by different cloud types.% We compare the cloud feedbacks calculated using the \citet{zelinka_computing_I_2012} cloud radiative kernels to the estimate of cloud feedback calculated using the radiative kernels of \citet{pendergrass_surface_2018} and \citet{huang_pattern_2017}. We find that the total cloud feedback is not sensitive to the method of calculation (not shown).

The kernels calculate SW and LW cloud feedbacks separately for various cloud-top pressure and cloud optical depth values. This allows for decomposition of the cloud feedback into SW and LW contributions and into the six cloud feedback components of \citet{sherwood_assessment_2020}, which are calculated in the CAM6 PPE using the methods of \citet{zelinka_evaluating_2022}. Obscuration effects, the hiding and revealing low clouds by changes in high clouds, are accounted for. Ascent and descent regions are defined by the sign of vertical velocity at 500 hPa at that location\add{ in each simulation}. The cutoff between high and low clouds is 680 hPa. The high-cloud altitude and land cloud amount feedbacks are calculated globally. The tropical marine low-cloud and tropical anvil cloud area feedbacks are calculated between 30$^\circ$S and 30$^\circ$N. The middle latitude low-cloud amount feedback is calculated between 30$^\circ$N and 60$^\circ$N and between 30$^\circ$S and 60$^\circ$S. The high latitude low-cloud optical depth feedback is calculated between 40$^\circ$N and 70$^\circ$N and between 40$^\circ$S and 70$^\circ$S. Amount, altitude, and optical depth feedbacks linearly sum to the total feedback, so the cloud feedback components are separate from one another, and sum to the total cloud feedback when the unassessed component is included.

\subsection{CMIP6, AMIP, and assessed values}

We compare total radiative feedbacks, total cloud feedbacks, and SW and LW cloud feedbacks in the CAM6 PPE to those in CMIP6 models. Feedbacks from coupled experiments are calculated using the CMIP6 pre-industrial control (`piControl') and abrupt CO$_2$ quadrupling (`abrupt\add{-}4xCO2') experiments; we refer to these as CMIP6. Feedbacks from atmosphere-only experiments are calculated using `amip' (CMIP DECK) and `amip\change{\_}{-}p4K' (CFMIP, uniform 4K warming) experiments for model years 1983 through 2008, referred to as AMIP\footnote{We use 1983 through 2008 values instead of the full 1979-2014 because that is what is used in Zelinka et al. (2022), which was done for comparison with observations.}. Values for the feedbacks are provided by Mark Zelinka's `cmip56\_forcing\_feedback\_ecs' (https://github.com/mzelinka/cmip56\_forcing\_feedback\_ecs) and `assessed\_cloud\_fbks' (https://github.com/mzelinka/assessed-cloud-fbks) GitHub repositories, as described by \citet{zelinka_causes_2020, zelinka_evaluating_2022}. The corresponding calculations are repeated for the CAM6 PPE following the methods of \citet{zelinka_causes_2020, zelinka_evaluating_2022}. CESM2 (CMIP) is not included in the six cloud feedback components of \citet{zelinka_evaluating_2022} because one of the needed variables (clisccp) is not available for the piControl simulation.

%\begin{table}[]
%    \centering
%    \begin{tabular}{c|c|c|c}
%        Kernel & Pendergrass & Huang & Zelinka \\
%         \hline
%        Kernel data & CAM5 (PORT) & ERA-Interim (RRTM) & CFMIP (Fu-Liou) \\
%        Feedbacks & Planck and LR (Temp), surface alb, WV, SW and LW cloud & & SW and LW cloud\\
%        Features & & & Decompose by optical depth and level \\
%    \end{tabular}
%    \caption{Caption}
%    \label{tab:my_label}
%\end{table}

\section{Spread in cloud feedbacks} \label{sec:spread}

\begin{figure}
    \centering
    \includegraphics[width = 300pt]{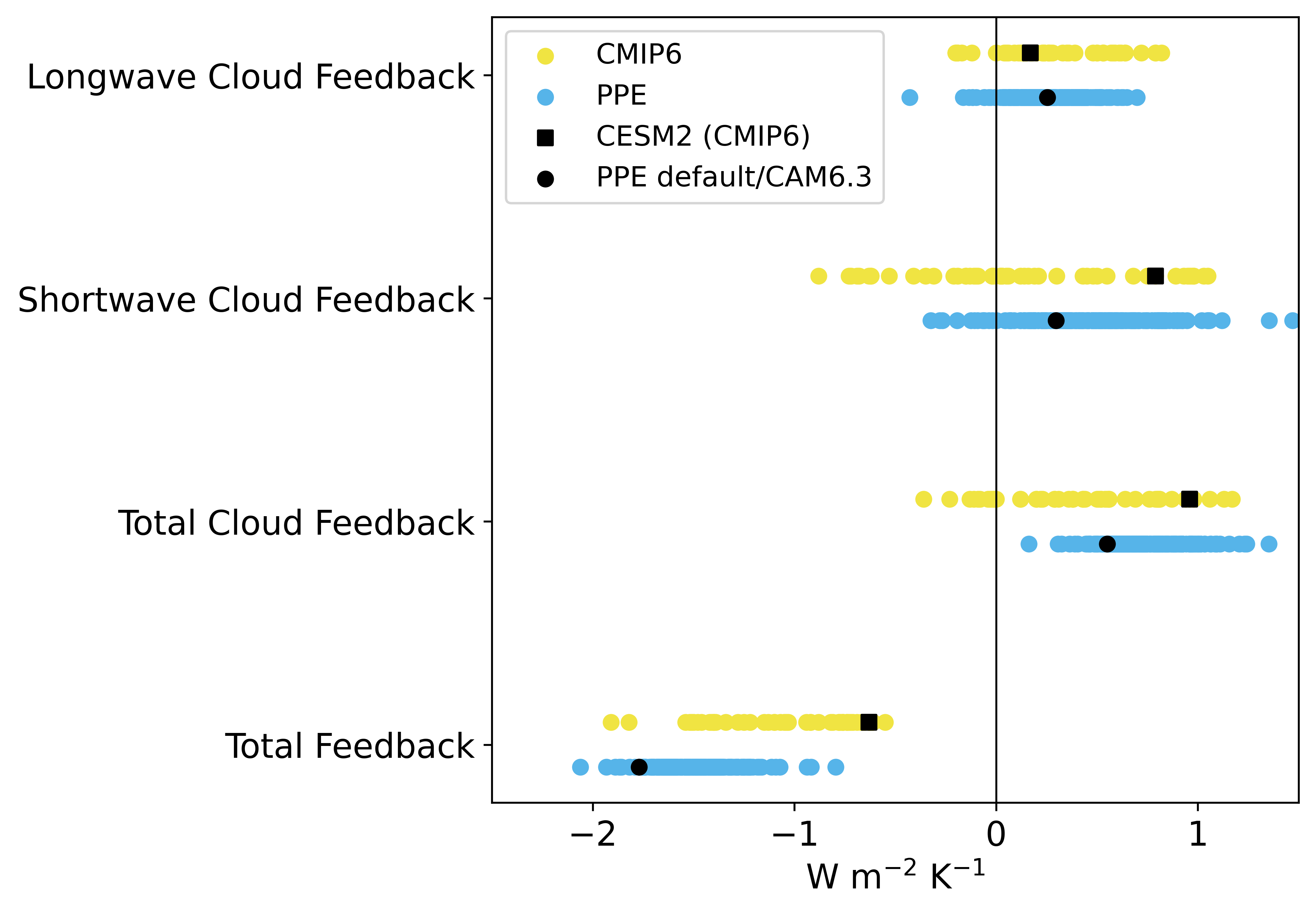}
    \caption{Spread in global-mean radiative feedback across CMIP6 models (yellow) and across PPE (blue). From bottom to top are the total radiative feedbacks, the cloud-radiative feedbacks, SW cloud-radiative feedbacks, and LW cloud-radiative feedbacks. The black square over the AMIP models denotes the CESM2 values and the black dot over the CAM6 PPE indicates the CAM6 PPE default simulation value.}
    \label{fig:PPE_CMIP_spread}
\end{figure}

We begin by comparing the distribution of total cloud radiative feedbacks between the subset of 206\add{ CAM6} PPE simulations and the CMIP6 models. \add{The spread across the CMIP6 models represents structural uncertainty; each model solves different equations with different numerical methods. On the other hand, the spread across the CAM6 PPE represents parametric uncertainty; the only difference between members is the parameter values. }The total feedback, along with the total cloud feedback and its SW and LW components, are shown in Figure \ref{fig:PPE_CMIP_spread}\remove{ (see Figure S1 for the ensemble mean and standard deviation)}. The\add{ CAM6} PPE default simulation has a smaller (more negative) total feedback and a smaller (less positive) total cloud feedback than the CESM2 coupled experiment (Figure \ref{fig:PPE_CMIP_spread}; compare black square and black circle). This is an important and unexpected finding which we discuss further in Section \ref{sec:low_cf}. We also find that the spread across total cloud feedbacks in the CAM6 PPE and CMIP6 models are comparable to one another. \remove{The CAM6 PPE has a larger ensemble-mean total cloud feedback than the CMIP6 models and a smaller standard deviation (Figure S1), which may be because the CAM6 PPE ensemble varies only in parameter values while CMIP6 models also have structural differences. }There is no \textit{a priori} expectation for the CAM6 PPE spread to be similar to the CMIP6 spread. The two ensembles are entirely different with different numbers of members and different sources of spread (i.e., systematic parameter changes in the CAM6 PPE and opportunistic and structural differences across CMIP6).

In order to compare the processes setting the comparable spreads in cloud feedbacks across the CAM6 PPE simulations and CMIP6 models, we further decompose cloud feedbacks into 1) their shortwave (SW) and longwave (LW) contributions and 2) into the six cloud feedback components of \citet{sherwood_assessment_2020}.\add{ Other decompositions are possible; we use the SW and LW decomposition because of its simplicity and the six cloud feedback components of }\citet{sherwood_assessment_2020}\add{ for comparison with recent work }(e.g. \cite{zelinka_evaluating_2022}). The SW and LW contributions to the cloud feedback are shown in Figure \ref{fig:PPE_CMIP_spread} and compare favorably to one another. A bar graph with the mean and standard deviation of these data are shown in Figure S1. \remove{We find that the larger mean cloud feedback in the CAM6 PPE is from a larger SW contribution.}

\begin{figure}
    \centering
    \includegraphics[width = \textwidth]{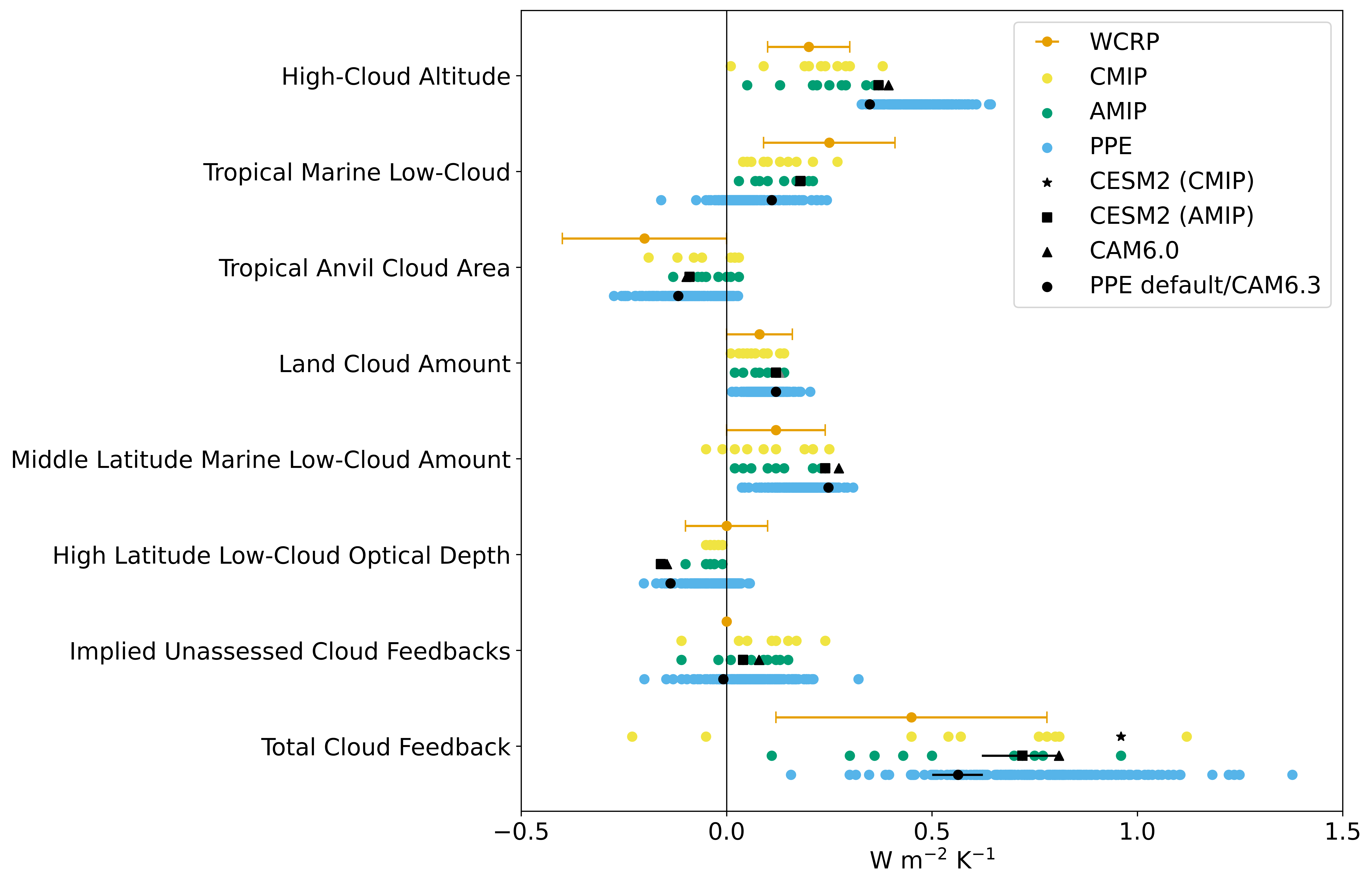}
    \caption{Comparison of total cloud feedbacks, unassessed cloud feedbacks, and cloud feedback components in WCRP assessment (orange), CMIP6 models (yellow), AMIP models (green), and the CAM6 PPE (blue). The range of the WCRP assessed feedbacks represents the 1$\sigma$ spread. The black star in the CMIP total cloud feedback row denotes the CESM2 (CMIP) value. The black squares in the AMIP rows denotes the CESM2 (AMIP) values and the bar going through it in the total cloud feedback row covers the interannual spread. The black triangle in the AMIP total feedback row denotes the simulation of CAM6.0 forced with F2000climo SST (see text for details). The black dots in the CAM6 PPE rows denote the CAM6 PPE default simulation values, which use CAM6.3 and F2000climo, and the black bar going through it in the total cloud feedback row covers the interannual spread in a 12-year version of the CAM6 PPE default simulation.}
    \label{fig:assessed}
\end{figure}
%[CESM2-CAM6.0, CESM2-CAM6.0 (AMIP), CAM6.0 (F2000 climo), PPE default CAM6.3 (F2000 climo)]

The total cloud and the six components of \citet{sherwood_assessment_2020} are shown in Figure \ref{fig:assessed} for the CMIP6 models, AMIP models, and CAM6 PPE. The 1$\sigma$ range of each component in the WCRP assessment by \citet{sherwood_assessment_2020} is also included for comparison. A bar graph with the mean and standard deviation of these data are shown in Figure S2. The AMIP models are more similar to the CAM6 PPE simulations than the CMIP6 models because the AMIP and CAM6 PPE simulations are all atmosphere-only with imposed uniform 4K warming. Therefore, we focus on the comparison between CAM6 PPE and AMIP simulations.

We begin by comparing the mean of the cloud feedback components between the CAM6 PPE and the AMIP models. The mean total cloud feedback is larger in the CAM6 PPE than in the AMIP models. This larger total cloud feedback is the consequence of much larger high-cloud altitude feedback, which is partially compensated by a smaller mean tropical marine low-cloud feedback. The other four feedbacks have mean values which fall within the assessed 1$\sigma$ range of \citet{sherwood_assessment_2020}.\add{ Additionally, the mean unassessed cloud feedback component is close to zero in the PPE, but this component is the least understood and it is not known whether or not it should be zero.} The AMIP models have very similar mean cloud feedbacks to the CMIP6 models.

We next compare the spread in the cloud feedbacks across the CAM6 PPE and AMIP models. The range (largest minus smallest value) in the total cloud feedback are comparable between AMIP (0.85 W m$^{-2}$ K$^{-1}$) and the CAM6 PPE (1.22 W m$^{-2}$ K$^{-1}$). This is reflected in each of the six cloud feedback components, which have similar ranges across the two ensembles. The AMIP models have very similar spreads in cloud feedbacks to the CMIP6 models. However, the CMIP6 range for the high latitude low-cloud optical depth feedback is much smaller than the \add{CAM6 }PPE range, and also much smaller than the assessed 1$\sigma$ range of \citet{sherwood_assessment_2020}.

Overall, the comparable spreads in total cloud feedback between the CAM6 PPE, CMIP6, and AMIP models is the result of comparable spreads in SW and LW cloud feedbacks and of comparable spreads in the six cloud feedback components. Further, no individual component of the cloud feedback is controlling the spread in any of the three ensembles, and these different feedback components contribute about the same to the total cloud feedback spread in the three ensembles. Again, we emphasize that there is no\add{ \emph{a priori}} reason to expect a single model's estimated parametric uncertainty to reproduce the spread in cloud feedbacks in CMIP6 or AMIP.\add{

Why might the parametric and structural uncertainties be comparable? Why does it matter? }\add{That the spreads in cloud feedbacks due to parametric and structural uncertainties are similar suggests that the processes controlling cloud feedbacks can be influenced by either structural or parametric differences. For example, if some cloud property is contributing to a feedback and that property varies with structural and parametric differences which are similarly uncertain, then the portion of the cloud feedbacks set by that property would have similar spreads across CMIP6 and the CAM6 PPE. The CAM6 PPE is advantageous because the only difference between ensemble members is parameter values and each parameter controls a process in the model. Therefore, in Section }\ref{sec:param}\add{ we evaluate the influence of various parameters on feedbacks.} \change{Also, the CMIP6 and AMIP ensembles have very similar total cloud feedbacks and cloud feedback components as one another. However, t}{

Notably, t}he total cloud feedback is smaller in the \add{CAM6 }PPE default simulation than the CESM2 (AMIP) simulation, which is smaller than the CESM2 (CMIP) simulation. We investigate these differences in \change{the following section.}{Section }\ref{sec:low_cf}.

\remove{Differences in total cloud feedback across CESM2 experiments}

\remove{From the `Total Cloud Feedback' row of Figure 3 it is clear that the CESM2 (CMIP) simulation has a larger total cloud feedback (0.96 W m$^{-2}$ K$^{-1}$) than the CESM2 (AMIP) simulation (0.72 W m$^{-2}$ K$^{-1}$), which is larger still than that of the PPE default simulation (0.56 W m$^{-2}$ K$^{-1}$). These discrepancies are quite large. To demonstrate that these discrepancies in cloud feedback are substantial, we solve for their corresponding ECS values using CESM2 (CMIP) values of $\Delta F$ and $\lambda_{noncloud}$. We find substantially different corresponding ECS values of 5.1 K in CESM2 (CMIP), 3.7 K in CESM2 (AMIP), and 3.1 K in CAM6 PPE default.}

\remove{The difference in total cloud feedback between CESM2 (CMIP) and CESM2 (AMIP) may be attributable to 1) different abilities of these simulations to represent the ``pattern" effect and/or 2) differences in high latitude optical depth feedbacks. In GCMs, the pattern effect describes a time evolution of radiative feedbacks, including cloud radiative feedbacks, which is largely attributable to the evolution of SST patterns over time }\remove{. The CESM2 (CMIP) cloud feedback is calculated using 150 year piControl and 4xCO$_2$ simulations. In contrast, the CESM2 (AMIP) cloud feedback is calculated using 25 year amip and amip\_p4K (i.e. uniform warming) simulations. Therefore, the pattern effect influences the cloud feedback in the CESM2 (CMIP) simulation, but is not represented by the CESM2 (AMIP) simulation because it has uniform 4K warming. We note that the pattern effect is large in CESM2 as compared to other models and an observation-based estimate}\remove{. Another possible explanation for the larger cloud feedback in CESM2 (CMIP) than CESM2 (AMIP) is the evolution of a high latitude optical depth feedback. This feedback is negative and goes to zero with warming, and is the result of decreases in ice clouds and increases in liquid cloud over the southern ocean with warming. However, as the planet warms there are fewer ice clouds and more liquid clouds, so this feedback approaches zero. This feedback increases from -1.25 W m$^{-2}$ K$^{-1}$ in the first 15 years to -0.02 W m$^{-2}$ K$^{-1}$ in the last 15 years of a 150-year fully-coupled 4xCO$_2$ simulation of CESM2 }\remove{. The value of this feedback in CESM2 (AMIP) is -0.97 W m$^{-2}$ K$^{-1}$. We do not have the necessary output to calculate this feedback in CESM2 (CMIP), but we hypothesize that it is likely larger (less negative) in the 150-year 4xCO$_2$ CESM2 (CMIP) simulation than in the CESM2 (AMIP) simulation. Therefore, this high latitude optical depth feedback is another potential explanation for the discrepancy in cloud feedbacks between the CESM2 (AMIP) and CESM2 (CMIP).}

\remove{On the other hand, the difference in total cloud feedback between the default CAM6 PPE simulation (0.56 W m$^{-2}$ K$^{-1}$) and the CESM2 (AMIP) simulation (0.72 W m$^{-2}$ K$^{-1}$) is surprising. Differences between the CESM2 (AMIP) and CAM6 PPE default simulations include 1) the CAM6 PPE default simulation is only 3 years long while the CESM2 (AMIP) simulation is 25 years long, 2) the CAM6 PPE default simulation is forced with different SSTs than the CESM2 (AMIP) simulation (the CAM6 PPE SSTs do not have interannual variability while CESM2 (AMIP) SSTs do), and 3) CESM2 (AMIP) uses CAM6.0 whereas the PPE default simulations use CAM6.3. In order to evaluate whether simulation length or interannual variability can explain the discrepancy in total cloud feedback, we calculate the interannual spread in total cloud feedback for each of the 25 years of the CESM2 (AMIP) simulation and for a 12 year version of the CAM6 PPE default simulation (Figure }\remove{. We find that neither spread explains the difference in total cloud feedback.}

\remove{In order to further evaluate which of these differences explains the discrepancy in total cloud feedback, we run a 3-year simulation using CAM6.0 (the version of CAM used in CESM2 (AMIP)) forced with F2000climo SSTs (the same SSTs as CAM6 PPE). This simulation, which we refer to as CAM6.0, differs from the PPE default (CAM6.3) simulation only in model generation. The total feedback in CAM6.0 (F2000climo) is 0.81 W m$^{-2}$ K$^{-1}$, which is larger than the CESM2 (AMIP) simulation but within the range of interannual variability (Figure (0.72 W m$^{-1}$), falling near the high end of interannual variability (compare the black square and its whiskers with the black triangle in Figure }\remove{. Therefore, }\remove{, it appears that modifications to CAM6 made between CAM6.0 and CAM6.3 have reduced the climate sensitivity. Parameter values did not change between these two experiments, so this difference is attributable entirely to structural modifications to CAM6.}

\remove{Having identified that structural changes between model versions impact the total cloud feedback, we next evaluate the difference between the six cloud feedback components in CAM6.0 and CAM6.3 (Figure }\remove{). Figure}\remove{ reveals that the difference in cloud feedback is due primarily to differences high-cloud altitude, tropical marine low-cloud, and unassessed cloud feedbacks. We further evaluate the difference between cloud feedbacks between CAM6.0 and CAM6.3 by plotting the high, mid-level, and low cloud contributors to both the SW and LW cloud feedbacks (Figure }\remove{). From Figure }\remove{, it is clear that the decrease in cloud feedback is primarily due to the SW contribution, with small decreases due to the LW contribution. Further, the SW contribution is dominated by the low-cloud contribution in the tropics and middle latitudes. To the extent that the LW cloud feedback decreases is mostly due to changes in the high-cloud contribution in the tropics and subtropics. We leave further evaluation of the difference in cloud feedback between CAM6.0 and CAM6.3 to future work.}

\section{Parameters}\label{sec:param}

\subsection{Which parameters set the spread in cloud feedbacks?}
\note{moved this up since I think it follows the previous section well.}
An advantage of using a perturbed parameter ensemble is the ability to identify the influence of various parameters on the cloud feedbacks.\add{ Parameters correspond to physical processes, so understanding which parameters control the spread in a given feedback offers some insights into the processes controlling that feedback in CAM6.} We use regression models to quantify the influence of each parameter on the spread in the total cloud feedback, each of its six components, and the unassessed cloud feedback. Scatterplots of each parameter against each of the cloud feedbacks studied here indicate that it is reasonable to assume linearity. The linear regression model is given by 
\begin{equation} \label{eq:4}
    \hat{\lambda_j} = a_{0,j}+\sum_{i=0}^{45}a_{i,j}\tilde{p_i},
\end{equation}
where $\hat{\lambda_j}$ is the estimated cloud feedback component, $a_{0,j}$ is the intercept for a cloud feedback component $j$, $a_{i,j}$ are the regression coefficients for each parameter $i$ and each cloud feedback component $j$, and $\tilde{p_i}$ are the parameter values which are scaled to fall between 0 and 1 according to
\begin{equation} \label{eq:5}
    \tilde{p_i} = \frac{p_i-p_{i,min}}{p_{i,max}-p_{i,min}},
\end{equation}
where $p_i$ are the unscaled parameter values, $p_{i,min}$ is the minimum value in $p_i$, and $p_{i,max}$ is the maximum value in $p_i$. In practice, we rescale the parameters using Python's sklean.preprocessing.MinMaxScaler function. In order to avoid overfitting due to the large number of parameters (43) relative to feedbacks (206), we use a \change{Lasso}{least absolute shrinkage and selection operator (LASSO)} regression, which imposes an L1 penalty to yield a sparse model (i.e. a model in which some parameters are eliminated by setting their coefficient to 0). The degree of sparseness is set by a tuning parameter. Here, the tuning parameter is calculated using an automated, 5-fold cross validation function (sklearn.linear\_model.LassoCV in Python with default function parameters). While it is common to use machine learning emulation where we use a \change{Lasso}{LASSO} regression (e.g. \citet{dagon_machine_2020}), we find using 20-member test datasets that a \change{Lasso}{LASSO} regression model has comparable mean-squared error to Gaussian process and random forest machine learning models for the cloud feedbacks studied here\change{ (see}{. See} \citet{watsonparris_climatebench_2022} for more details about machine learning emulators\remove{)}.

The six cloud feedback components are separate from one another, and approximately sum to the total cloud feedback. Therefore, the sum of the regression models given by Equation \ref{eq:4} for each of the six cloud feedback components yields an estimate of the total cloud feedback, as follows
\begin{equation} \label{eq:6}
    \lambda_{cloud}\approx\sum_ja_{0,j}+\sum_{i=0}^{45}\left(\sum_ja_{i,j}\right)\tilde{p_i},
\end{equation}
where $j$ are the cloud feedback components. Conveniently, the regression coefficients $a_{i,j}$ can be interpreted as the spread of estimated cloud feedback across the CAM6 PPE due to variations in the corresponding parameter. This nice property exists because the parameter values $\tilde{p_i}$ are scaled to fall between 0 and 1. More specifically, a given parameter's influence on cloud feedback component $j$ is estimated by $a_{i,j}\tilde{p_i}$ for some parameter with index $i$. Since the smallest $\tilde{p_i}$ value is always 0 and the largest $\tilde{p_i}$ value is always 1, the difference in $\lambda_j$\add{ is} due to variations in parameter $\tilde{p_i}$ is $a_{i,j}$. For example, if a parameter has a regression coefficient of 0.1, all else equal, the estimated cloud feedback component will be 0.1 W m$^{-2}$ K$^{-1}$ greater when the parameter value is 1 than when it is 0. Therefore, a regression coefficient of 0.1 means that variations in that parameter are contributing a spread of 0.1 W m$^{-2}$ K$^{-1}$ to the estimated cloud feedback component. Further, the contribution of a given parameter's variations on total cloud feedback is approximately equal to $\sum_{j}a_{i,j}\tilde{p_i}$. In short, larger regression coefficients indicate more contribution to the spread across the CAM6 PPE.

Table \ref{tab:assessed} shows the regression coefficients for each cloud feedback component for the three parameters most highly correlated with total cloud feedback: \texttt{CLUBB\_C8}, \texttt{MG2\_DCS}, and \texttt{ZM\_capelmt}. \texttt{CLUBB\_C8} is a skewness coefficient associated with the third moment of vertical velocity. Larger \texttt{CLUBB\_C8} values correspond to thicker, more reflective clouds. \texttt{MG2\_DCS} is the ice-snow autoconversion size threshold; a larger value favors more ice crystals over snow in the atmosphere. \texttt{ZM\_capelmt} is a CAPE threshold value for triggering deep convection. \add{Therefore, these processes have the largest influence on the total cloud feedback in the CAM6 PPE. }Table S1 shows the regression coefficients for each cloud feedback component for all 43 parameters. \change{This approach}{Comparing the regression coefficients} allows us to see 1) how much spread in the total feedback is associated with each parameter, and 2) how each of the six cloud feedback components contributes to the total spread. More specifically, according to this simple model, about 0.14 W m$^{-2}$ K$^{-1}$ of the spread in total cloud feedback is attributed to variations in \texttt{CLUBB\_C8}, and that about 0.06 W m$^{-2}$ K$^{-1}$ of that spread is from the tropical marine low-cloud feedback. \change{Further, a}{A}bout 0.25 W m$^{-2}$ K$^{-1}$ of the spread in total cloud feedback is attributed to variations in \remove{the }\texttt{MG2\_DCS}\remove{ parameter}, and that about 0.09 W m$^{-2}$ K$^{-1}$ of that spread is from the tropical anvil cloud area feedback and about 0.05 W m$^{-2}$ K$^{-1}$ of that spread is from the high-cloud altitude feedback. Further, \remove{the }\texttt{ZM\_capelmt}\remove{ parameter} contributes about 0.17 W m$^{-2}$ K$^{-1}$ of the spread in total cloud feedback, and about 0.08 W m$^{-2}$ K$^{-1}$ is from the tropical anvil cloud area feedback and about 0.07 W m$^{-2}$ K$^{-1}$ is from the high latitude low-cloud optical depth feedback. The sum of the six cloud feedback component regression coefficients may differ from the total cloud feedback regression coefficient because of 1) unassessed cloud feedback, and 2) statistical model errors. We also show the regression coefficients for the unassessed cloud feedback, and they partially explain the discrepancy for many of the parameters (Tables \ref{tab:assessed} and S1).

The simple \change{Lasso}{LASSO} regression approach is very useful for estimating the influence of each parameter on cloud feedbacks, but is subject to some limitations. The regression model does not capture nonlinear relationships between parameter and feedback nor the influence of interactions between parameters on feedback.\add{ Nonetheless, given the comparable spreads in cloud feedbacks between CMIP models and the CAM6 PPE, the processes associated with these three influential parameters may be promising areas of future model development in order to constrain cloud feedbacks.}

\begin{sidewaystable}
    \caption{LASSO regression coefficient for each parameter, multiplied by 100 by visual clarity.}
    \begin{tabular}{c|c|c|c|c| c c c c c c|c}
    \textbf{Scheme} & \textbf{Parameter name} & \textbf{Description} & \textbf{Total} & \textbf{Sum assessed} & \textbf{Hi Alt} & \textbf{Trop Lo} & \textbf{Trop Anvil} & \textbf{Land} & \textbf{Mid Lo} & \textbf{Hi Lo} & \textbf{Unassessed}\\

    \hline
    CLUBB & C8 & Coef. \#1 in C8 skewness Eqn & 14.04 & 9.71 & -0.74 & 6.35 & 0 & 1.66 & 1.54 & 0.91 & 3.03\\
    \hline
    MG2/PUMAS & DCS & Autoconversion size threshold ice-snow & 25.01 & 16.33 & 4.72 & 0.31 & 8.72 & 0.96 & 2.64 & -1.02 & 8.12\\
%    & accre enhan fact & & 7.90 & 3.94 & -0.94 & 0 & 4.20 & -4.28 & -1.22 & 6.18 & 5.59\\
    \hline
    ZM & capelmt & Triggering threshold for ZM convection & 16.68 & 14.45 & 0.23 & -0.74 & 7.82 & -1.15 & 0.96 & 7.33 & 1.95\\
    \end{tabular} 
    \label{tab:assessed}
\end{sidewaystable}

\subsection{Influence of parameters on mean-state cloud errors and total cloud feedback}

\begin{figure}[h]
    \centering
    \includegraphics[width = \textwidth]{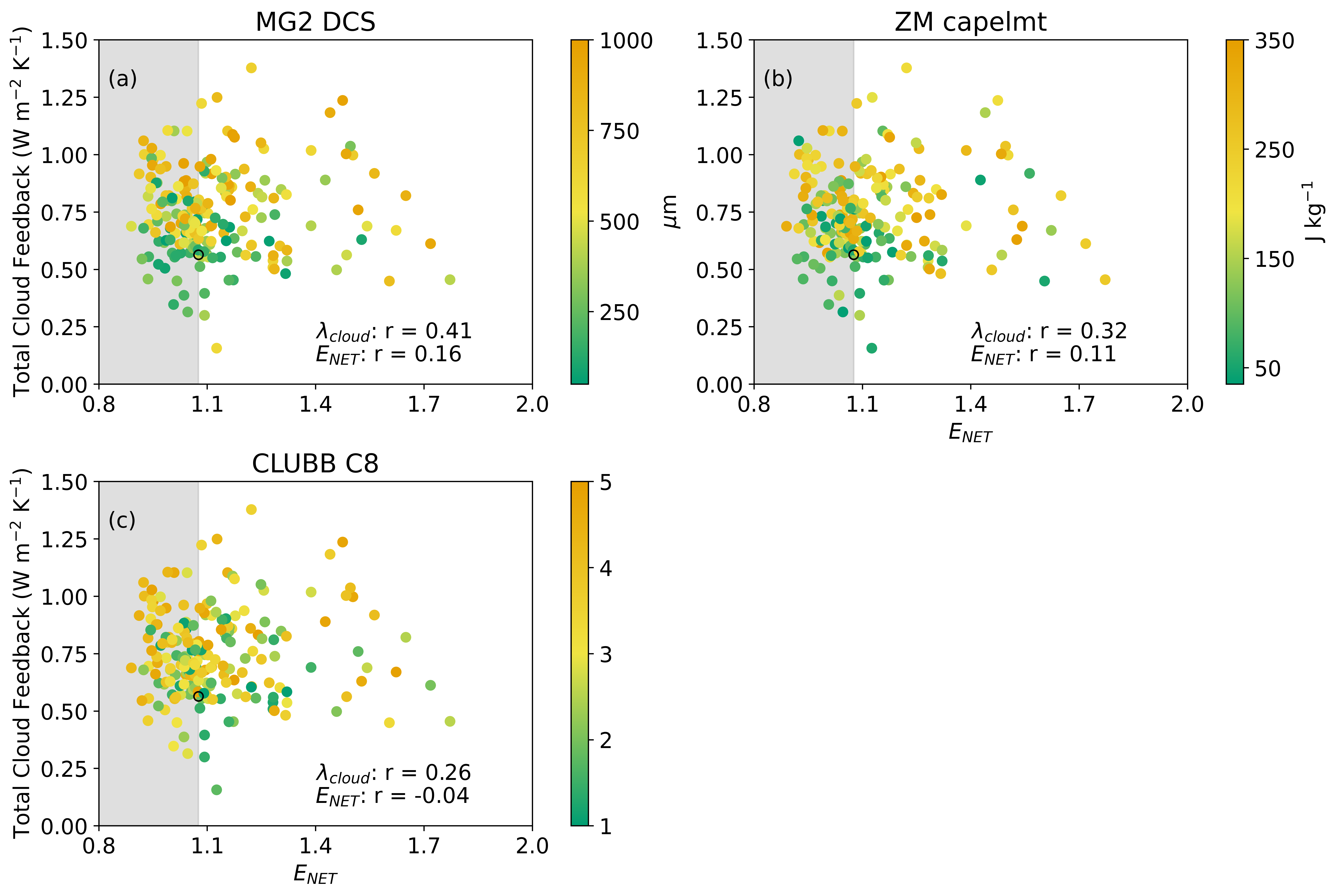}
    \caption{Scatterplot of total cloud feedback versus cloud feedback E$_{NET}$ in the CAM6 PPE with color of dots corresponding to the parameter value for three parameters. Each panel corresponds to a different parameter; the three most influential parameters with respect to \textit{cloud feedback} are represented. Grey shading covers the region where E$_{NET}$ values are equal to or smaller than that of the default simulation. The default simulation is encircled in black.}
    \label{fig:RMSE_param}
\end{figure}

\begin{figure}
    \centering
    \includegraphics[width = \textwidth]{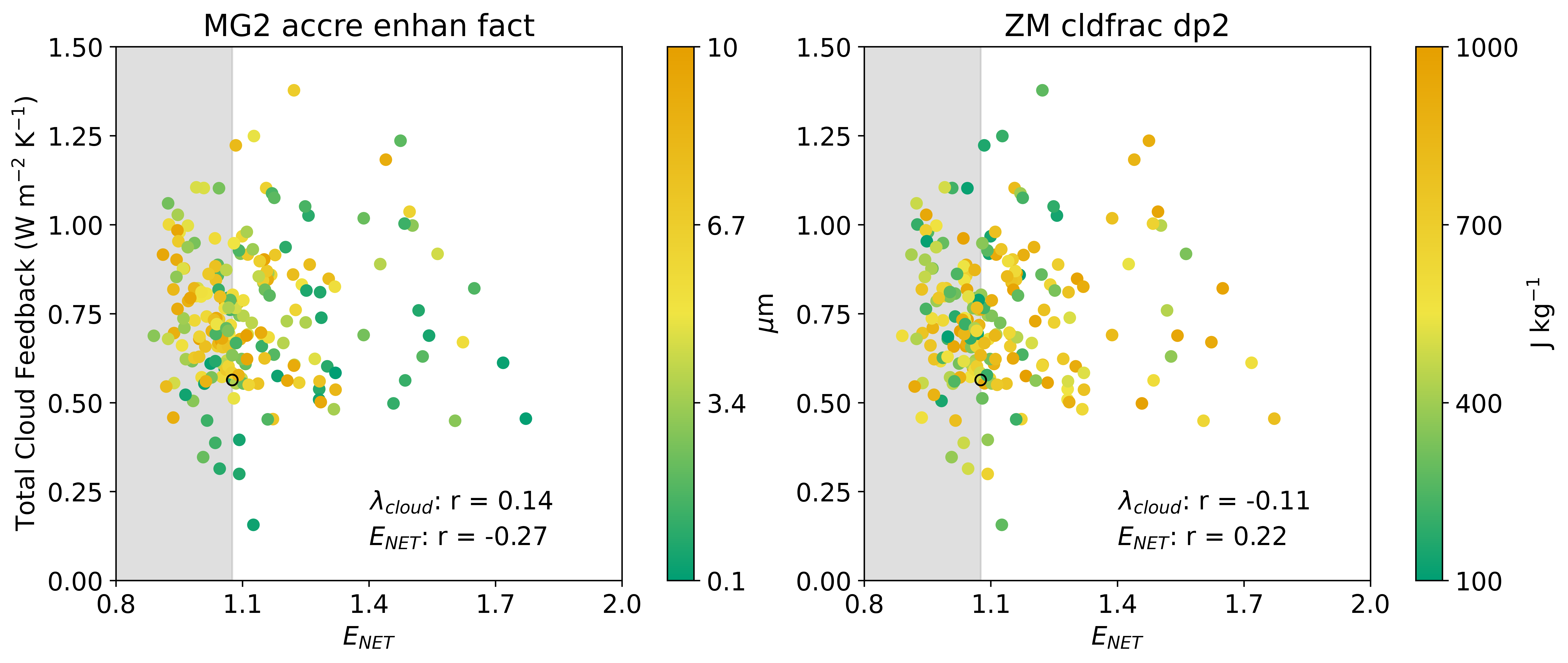}
    \caption{Same as Figure \ref{fig:RMSE_param} except the two most influential parameters with respect to E$_{NET}$ are shown.}
    \label{fig:ENET_param}
\end{figure}

\add{We plot each parameter as a function of mean-state cloud error and feedbacks. We evaluate the total cloud feedback as compared with 1) mean-state cloud errors and 2) the WCRP assessed estimate of the feedback components. We use E$_{NET}$ to evaluate mean-state cloud errors and an aggregated difference metric, RMSD, to compare with the WCRP estimate. The E$_{NET}$ metric is described in Section 2a and in }\citet{klein_are_2013} \add{and }\citet{zelinka_evaluating_2022}\add{. Following }\citet{zelinka_evaluating_2022}\add{, the difference metric used to compare with WCRP estimates is the root-mean-square deviation (RMSD) of the estimated feedback as compared to the central WCRP assessed value for each of the six assessed feedback components. The aggregated metric is simply the square root of the average of the six squared differences. Therefore, the difference metric is minimized when all six of the assessed cloud feedbacks are close to their assessed value. }\remove{In order to visualize the influence of each parameter on mean-state cloud errors (E$_{NET}$) and total cloud feedback, }\change{we elaborate on Figure 6(b) by adding color to show the values of}{We plot the parameter value as a function of mean state cloud error $E_{NET}$ and an aggregated distance metric for feedbacks RMSD for} the three parameters most highly correlated with cloud feedback: \texttt{MG2\_DCS}, \texttt{ZM\_capelmt}, and \texttt{CLUBB\_C8} (Figure \ref{fig:RMSE_param}), and of the two parameters most highly correlated with E$_{NET}$: \texttt{MG2\_accre\_enhan\_fact} and \texttt{ZM\_cldfrac\_dp2} (Figure \ref{fig:ENET_param}). These parameters were selected because their correlation coefficients are greater than 0.2. \texttt{MG2\_accre\_enhan\_fact} is an accretion enhancing factor, with larger values corresponding to more accretion of cloud droplets onto rain. \texttt{ZM\_cldfrac\_dp2} is a parameter for deep convective cloud fraction as a function of convective mass flux, with larger values corresponding to a higher deep convective cloud fraction. The equivalent figures for the remaining parameters are shown in Figures S3 through S7. The parameters most correlated with total cloud feedback are different from the parameters most correlated with E$_{NET}$, which is not surprising given the lack of correlation between E$_{NET}$ and total cloud feedback (\add{Section }\ref{sec:RMSD}\add{ and }Figure \ref{fig:RMSE}(\change{b}{c})).

\section{CAM6 PPE constraint on total cloud feedback} \label{sec:RMSD}

Given the large parametric sensitivity of cloud feedbacks in the CAM6 PPE, we ask whether the CAM6 PPE provide a constraint on the total cloud feedback. \add{To that end, we further compare the CAM6 PPE to mean-state cloud errors using $E_{NET}$ and to WCRP estimates of cloud feedbacks using RMSD.}\remove{To that end, we evaluate the total cloud feedback as compared with 1) mean-state cloud errors and 2) the WCRP assessed estimate of the feedback components. We use E$_{NET}$ to evaluate mean-state cloud errors and an aggregated difference metric, RMSD, to compare with the WCRP estimate. The E$_{NET}$ metric is described in Section 2a and in }\remove{, the difference metric used to compare with WCRP estimates is the root-mean-square deviation (RMSD) of the estimated feedback as compared to the central WCRP assessed value for each of the six assessed feedback components. The aggregated metric is simply the square root of the average of the six squared differences. Therefore, the difference metric is minimized when all six of the assessed cloud feedbacks are close to their assessed value.} The six assessed feedbacks nearly sum to the total cloud feedback, so the total cloud feedback must approach the assessed value as the RMSD approaches 0 unless there is a large unassessed component of the feedback. Importantly, it should be kept in mind that the assessed cloud feedback values are only estimates; they do not represent the true cloud feedbacks.\add{ In fact, several of the assessed values have been updated by more recent studies }\citep[e.g.][]{wall_observational_2022, mckim_physical_2023}. This difference metric is only as valuable as the assessed value. Another limitation of this approach is that the difference metric only uses the mean of the assessed range and neglects the estimated spread associated with each of the six cloud feedback components. For \change{this reason}{these reasons}, we put more emphasis on the constraint made using mean-state cloud errors than on the constraint made by comparing with WCRP values.

\begin{figure}
    \centering
    \includegraphics[width = \textwidth]{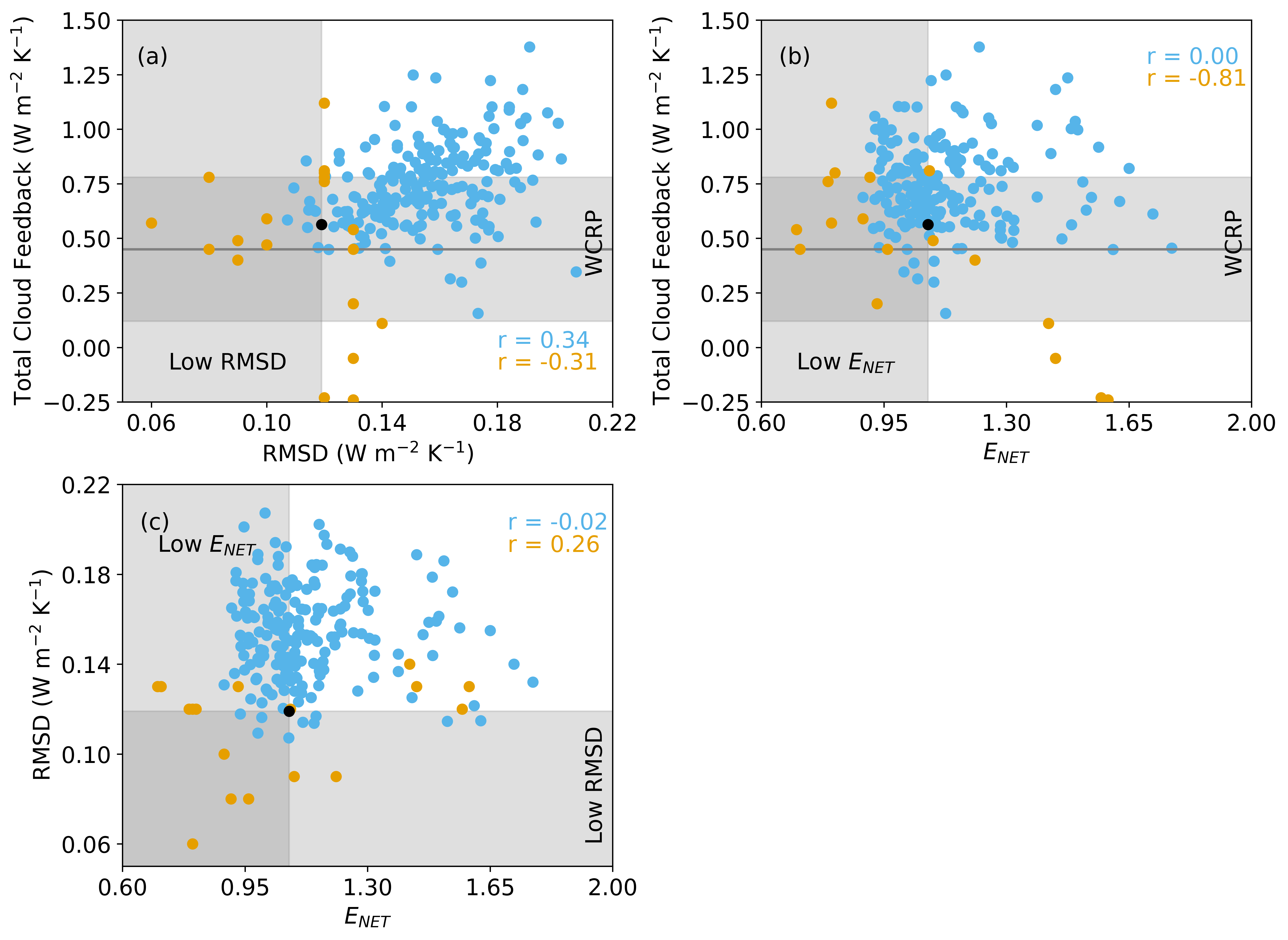}
    \caption{Scatterplot of total cloud feedback versus cloud feedback RMSD (a) and E$_{NET}$ (b) in the CAM6 PPE simulations (blue) and the CMIP5 and CMIP6 models (orange). The black dot denotes the default PPE simulation. In panels (a) and (b), the dark grey horizontal line is the WCRP assessed cloud feedback value and the horizontal grey shaded region covers the 1$\sigma$ WCRP range. The vertical grey shaded region covers the region where RMSD (a) and E$_{NET}$ (b) values are equal to or smaller than that of the default PPE simulation. Panel (c) shows a scatterplot of RMSD versus E$_{NET}$. The horizontal grey shaded region covers RMSD values equal to or smaller than that of the CAM6 PPE default simulation and the vertical grey shaded region covers E$_{NET}$ values equal to or smaller than that of the CAM6 PPE default simulation.\note{yellow to orange}}
    \label{fig:RMSE}
\end{figure}

We plot the total cloud feedback as a function of RMSD in Figure \ref{fig:RMSE}(a) and as a function of E$_{NET}$ in Figure \ref{fig:RMSE}(b). For reference, the CMIP5 and CMIP6 values are also included in the plot, and are the same as those from Figures 3 and 4 of \citet{zelinka_evaluating_2022}. Looking at RMSD in Figure \ref{fig:RMSE}(a), we find that the CAM6 PPE simulations exhibit some similar behavior to the CMIP models: simulations with a small RMSD tend to have more moderate cloud feedbacks than those with large RMSD. The large-RMSD models include both anomalously small and large cloud feedbacks. However, there are a number of differences in the behavior between the CAM6 PPE and the CMIP models of \citet{zelinka_evaluating_2022}. The CAM6 PPE ensemble members tend to have larger RMSD values than the CMIP models. The smallest RMSD value among the CAM6 PPE ensemble members is about 0.11 W m$^{-2}$ K$^{-1}$, whereas there are seven CMIP models with smaller RMSD values. Further, the distribution of total cloud feedbacks is shifted toward larger values in the CAM6 PPE as compared to CMIP models. Looking at Figure \ref{fig:assessed}, both of these difference\add{s} may be the result of comparatively large high-cloud altitude feedbacks in the CAM6 PPE.

Looking at Figure \ref{fig:RMSE}(b), the mean-state cloud errors tell a different story than that the \change{WCRP estimated values}{RMSD}. \change{The CMIP models are anticorrelated with E$_{NET}$}{In the CMIP models, total cloud feedback is anticorrelated with E$_{NET}$}, suggesting E$_{NET}$ constrains cloud feedbacks to be large. On the other hand, \add{in }the CAM6 PPE\add{,} \change{ensemble members are}{total cloud feedback is} uncorrelated with E$_{NET}$ and there is a large spread in cloud feedbacks amongst the small E$_{NET}$ ensemble members, suggesting that mean-state cloud errors do not provide an effective constraint on the total cloud feedback in the CAM6 PPE. Figure \ref{fig:RMSE}(c) plots the relationship between mean-state cloud errors (E$_{NET}$) and WCRP difference metric (RMSD), and we find that, consistent with the CMIP models, there is no relationship between the two metrics in the CAM6 PPE.

We conclude that we are unable to robustly constrain the total cloud feedback using the CAM6 PPE with available information. We leave further understanding of the lack of relationship between E$_{NET}$ and RMSD to future work. We also leave further understanding of the inconsistent relationship between E$_{NET}$ and cloud feedback across the CAM6 PPE and CMIP6 ensembles to future work.

\section{CAM5 to CAM6} \label{sec:CAM56}

Given the substantial influence of parameters on cloud feedbacks in CAM6, and the large increase in ECS from CAM5 to CAM6, we next evaluate whether changes in parameter values (parameter tuning) between CAM5 and CAM6 contribute to the increase in total cloud feedback and ECS \citep{gettelman_high_2019}.\note{ Gettelman citation added.}

\begin{figure}[h]
    \centering
    \includegraphics[width = 300pt]{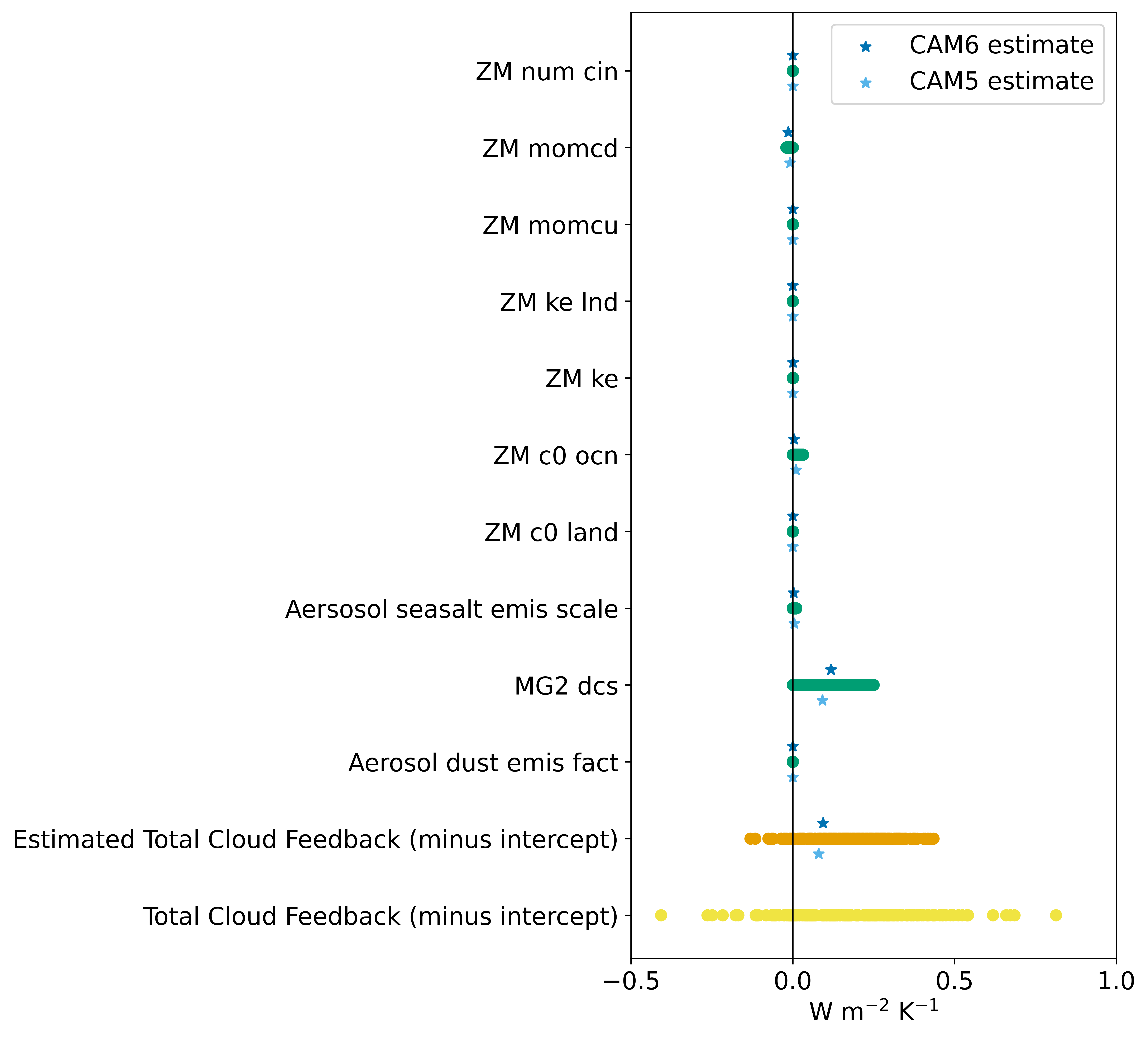}
    \caption{Total cloud feedback in each PPE simulation (yellow), LASSO regression estimate of total cloud feedback (orange), and the contribution to that estimate from the 10 parameters which 1) appear in both CAM5 and CAM6, and 2) changed from CAM5 to CAM6 (green). Stars denote the estimated total cloud feedback and the contributions from each parameter using CAM5 (light blue, below) and CAM6 (dark blue, above) parameter values. The regression intercept has been subtracted from both the estimated and actual cloud feedback values.}
    \label{fig:CAM56}
\end{figure}

To address this question, we develop a simple linear regression model of total cloud feedback as a function of each parameter value. We again use a \change{lasso}{LASSO} regression to avoid overfitting. The regression is similar to those of Equations \ref{eq:4} and \ref{eq:6}, except here the total cloud feedback is estimated directly instead of as the sum of the cloud feedback components. The simple regression model has a root-mean-square error (RMSE) of 0.13 W m$^{-2}$ K$^{-2}$. Figure \ref{fig:CAM56} compares the regression model estimate to the actual cloud feedback (minus the intercept). Looking at Figure \ref{fig:CAM56}, it is clear that this simple approach struggles to capture the very large and very small total cloud feedback values. Still, the RMSE value and Figure \ref{fig:CAM56} provide enough confidence in our simple model estimate of the total feedback to proceed.

In order to use this simple model to evaluate the role of parameter values in changes in total cloud feedback between CAM5 and CAM6, we create `CAM5' and `CAM6' parameter sets. The CAM6 parameter set is the default parameter set. On the other hand, the CAM5 parameter set uses the CAM5 values for the 10 parameters which 1) appear in both CAM5 and CAM6, and 2) change from CAM5 to CAM6 and default parameter values for the remaining parameters. Each parameter set is scaled to fall between 0 and 1 according to Equation \ref{eq:5}. Figure \ref{fig:CAM56} shows the contribution of each of these 10 parameters to the estimated total cloud feedback, which is the parameter value multiplied by its corresponding regression coefficient. Total cloud feedback values are then estimated by plugging the CAM5 and CAM6 parameter sets into the regression model. The CAM5 and CAM6 estimates and their contributions from each parameter is shown in Figure \ref{fig:CAM56}.

Examining Figure \ref{fig:CAM56}, we learn that there is negligible change in the estimated total cloud feedback due to differences in parameters from CAM5 to CAM6: only a 0.01 W m$^{-2}$ K$^{-1}$ increase. For reference, \citet{gettelman_high_2019} found an increase of 0.27 W m$^{-2}$ K$^{-1}$ from CAM5 to CAM6. This analysis points to structural changes to the model as the cause of the increase in cloud feedbacks\add{, not parameter tuning}.\remove{ This is consistent with the findings of Section 7, which found that changes in CAM6 between CAM6.0 and CAM6.3 decreased the cloud feedback without changing parameters.}

\section{Differences in total cloud feedback across CESM2 experiments} \label{sec:low_cf}

\add{From the `Total Cloud Feedback' row of Figure 3 it is clear that the CESM2 (CMIP) simulation has a larger total cloud feedback (0.96 W m$^{-2}$ K$^{-1}$) than the CESM2 (AMIP) simulation (0.72 W m$^{-2}$ K$^{-1}$), which is larger still than that of the CAM6 PPE default simulation (0.56 W m$^{-2}$ K$^{-1}$)}\footnote{The cloud feedback in amip-future4K is 0.50 W m$^{-2}$ K$^{-1}$, lower than any of the three simulations we focus on in this paper. This may be the result of very different SST patterns, especially in the tropical Pacific.}\add{. These discrepancies are quite large. To demonstrate that these discrepancies in cloud feedback are substantial, we solve for their corresponding ECS values using CESM2 (CMIP) values of $\Delta F$ and $\lambda_{noncloud}$. We find substantially different corresponding ECS values of 5.1 K in CESM2 (CMIP), 3.7 K in CESM2 (AMIP), and 3.1 K in CAM6 PPE default.}

\add{The difference in total cloud feedback between CESM2 (CMIP) and CESM2 (AMIP) may be attributable, not exhaustively, to 1) different abilities of these simulations to represent the ``pattern" effect and/or 2) differences in high latitude optical depth feedbacks. In GCMs, the pattern effect describes a time evolution of radiative feedbacks, including cloud radiative feedbacks, which is largely attributable to the evolution of SST patterns over time }\citep{dong_attributing_2019}.\add{ The CESM2 (CMIP) cloud feedback is calculated using 150 year piControl and abrupt-4xCO$_2$ simulations. In contrast, the CESM2 (AMIP) cloud feedback is calculated using 25 year amip and amip\_p4K (i.e. uniform warming) simulations. Therefore, the pattern effect influences the cloud feedback in the CESM2 (CMIP) simulation, but is not represented by the CESM2 (AMIP) simulation because it has uniform 4K warming. We note that the pattern effect is large in CESM2 as compared to other models and an observation-based estimate }\citep{andrews_effect_2022}\add{. Another possible explanation for the larger cloud feedback in CESM2 (CMIP) than CESM2 (AMIP) is the evolution of a high latitude optical depth feedback. This feedback is negative and goes to zero with warming, and is the result of decreases in ice clouds and increases in liquid cloud over the southern ocean with warming. However, as the planet warms there are fewer ice clouds and more liquid clouds, so this feedback approaches zero. This feedback increases from -1.25 W m$^{-2}$ K$^{-1}$ in the first 15 years to -0.02 W m$^{-2}$ K$^{-1}$ in the last 15 years of a 150-year fully-coupled abrupt-4xCO$_2$ simulation of CESM2 }\citep{bjordal_equilibrium_2020}\footnote{Bjordal et al. (2020) calculate the high latitude low-cloud optical depth feedback slightly differently differently from Sherwood et al. (2020) and elsewhere in this paper. Further, the details of the CESM2 simulations of Bjordal et al. (2020) are slightly different from those of CESM2 (CMIP).}.\add{ The value of this feedback in CESM2 (AMIP) is -0.97 W m$^{-2}$ K$^{-1}$. We do not have the necessary output to calculate this feedback in CESM2 (CMIP), but we hypothesize that it is likely larger (less negative) in the 150-year abrupt-4xCO$_2$ CESM2 (CMIP) simulation than in the CESM2 (AMIP) simulation. Therefore, this high latitude optical depth feedback is another potential explanation for the discrepancy in cloud feedbacks between the CESM2 (AMIP) and CESM2 (CMIP).}

\add{On the other hand, the difference in total cloud feedback between the default CAM6 PPE simulation (0.56 W m$^{-2}$ K$^{-1}$) and the CESM2 (AMIP) simulation (0.72 W m$^{-2}$ K$^{-1}$) is surprising. Differences between the CESM2 (AMIP) and CAM6 PPE default simulations include 1) the CAM6 PPE default simulation is only 3 years long while the CESM2 (AMIP) simulation is 25 years long, 2) the CAM6 PPE default simulation is forced with different SSTs than the CESM2 (AMIP) simulation (CAM6 PPE SSTs do not have interannual variability while CESM2 (AMIP) SSTs do), and 3) CESM2 (AMIP) uses CAM6.0 whereas the CAM6 PPE default simulations use CAM6.3}\footnote{More specifically, CESM2 (AMIP) uses cam6-0-026 and the CAM6 PPE are run with cam6-3-026.}.\add{ In order to evaluate whether simulation length or interannual variability can explain the discrepancy in total cloud feedback, we calculate the interannual spread in total cloud feedback for each of the 25 years of the CESM2 (AMIP) simulation and for a 12 year version of the CAM6 PPE default simulation (Figure }\ref{fig:assessed})\add{. We find that neither spread explains the difference in total cloud feedback.}

\add{In order to further evaluate which of these differences explains the discrepancy in total cloud feedback, we run a 3-year simulation using CAM6.0 (the version of CAM used in CESM2 (AMIP)) forced with F2000climo SSTs (the same SSTs as CAM6 PPE). This simulation, which we refer to as CAM6.0, differs from the CAM6 PPE default (CAM6.3) simulation only in model generation. The total feedback in CAM6.0 (F2000climo) is 0.81 W m$^{-2}$ K$^{-1}$, which is }\add{even }\add{larger than the CESM2 (AMIP) simulation}\change{ but within the range of interannual variability (Figure }{(0.72 W m$^{-1}$), falling near the high end of interannual variability (compare the black square and its whiskers with the black triangle in Figure }\ref{fig:assessed}).\add{ Note that the difference between the CESM2 (AMIP) simulation and CAM6.0 (F2000climo) simulation is likely attributable to the difference in SST between the two. The AMIP SSTs show a strong pattern effect over the duration of the historical period which is not captured by F2000climo, which is cyclic }\citep{andrews_effect_2022}. \change{Therefore}{Finally}, \add{by comparing the black circle with the black triangle in Figure }\ref{fig:assessed}\add{, it appears that modifications to CAM6 made between CAM6.0 and CAM6.3 have reduced the climate sensitivity. Parameter values did not change between these two experiments, so this difference is attributable entirely to structural modifications to CAM6.}

\begin{figure}[h]
    \centering
    \includegraphics[width = 250pt]{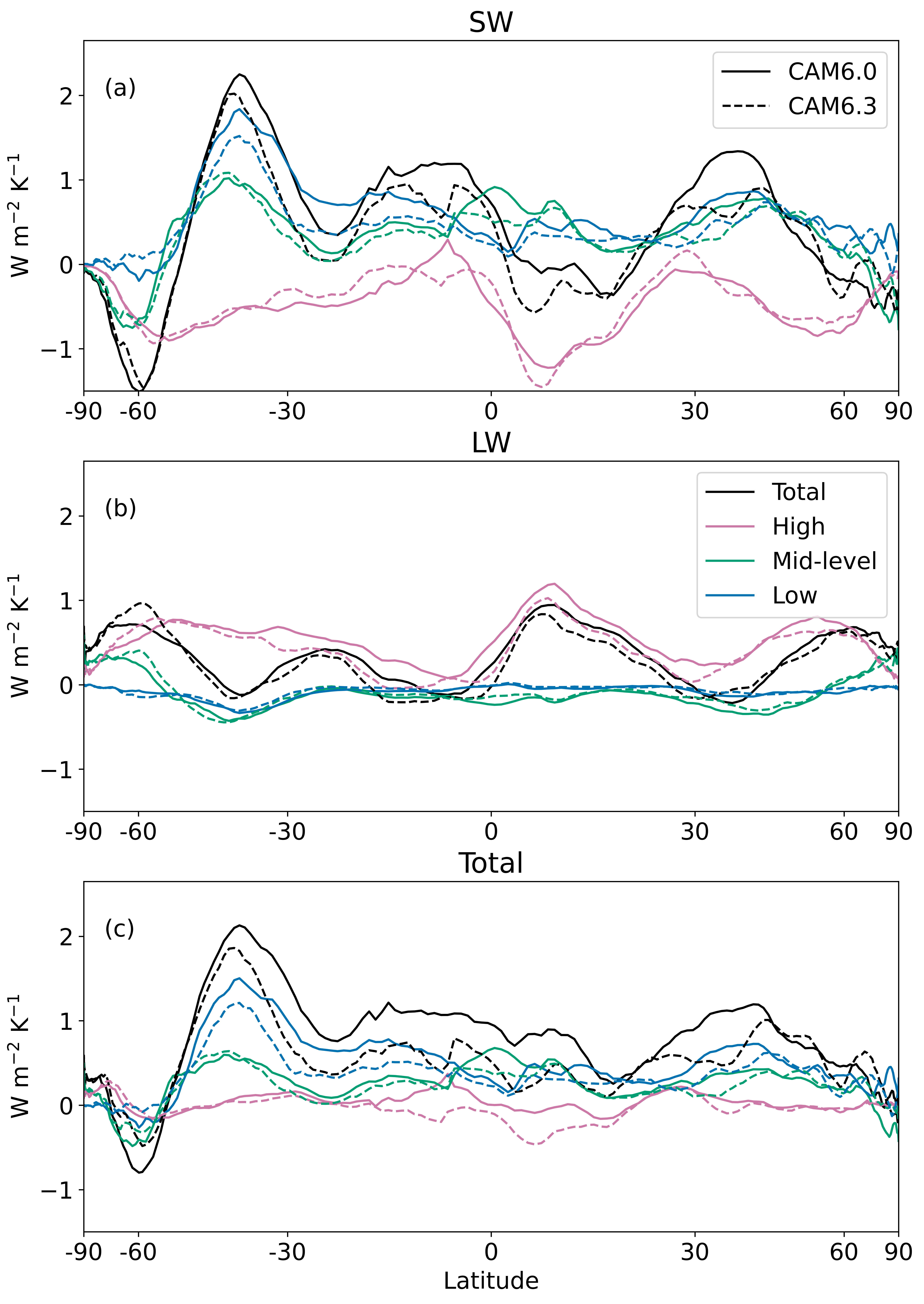}
    \caption{Zonal-mean cloud feedbacks in CAM6.0 (solid) and in CAM6.3 (dashed). The total cloud feedback (c) is decomposed into SW (a) and LW (b) contributions. The total cloud feedback (black) is also decomposed into high (\change{yellow}{pink}), mid-level (green), and low (blue) cloud contributions. The $x$ axis is scaled as the sine of latitude.\note{Colors have changed since initial submission.}}
    \label{fig:tag_de}
\end{figure}

\add{Having identified that structural changes between model versions impact the total cloud feedback, we next evaluate the difference between the six cloud feedback components in CAM6.0 and CAM6.3 (Figure }\ref{fig:assessed})\add{. Figure }\ref{fig:assessed}\add{ reveals that the difference in cloud feedback is due primarily to differences high-cloud altitude, tropical marine low-cloud, and unassessed cloud feedbacks. We further evaluate the difference between cloud feedbacks between CAM6.0 and CAM6.3 by plotting the high, mid-level, and low cloud contributors to both the SW and LW cloud feedbacks (Figure }\ref{fig:tag_de})\add{. From Figure }\ref{fig:tag_de}\add{, it is clear that the decrease in cloud feedback is primarily due to the SW contribution, with small decreases due to the LW contribution. Further, the SW contribution is dominated by the low-cloud contribution in the tropics and middle latitudes. To the extent that the LW cloud feedback decreases is mostly due to changes in the high-cloud contribution in the tropics and subtropics. We leave further evaluation of the difference in cloud feedback between CAM6.0 and CAM6.3 to future work.}

\section{Comparison with PaleoCalibr} \label{sec:paleo}

\add{To further understand our results, we compare the CAM6 PPE used here to another PPE which used the Paleoclimate-calibrated CAM6 (CAM6-PaleoCalibr), which has a smaller ECS than the standard version of CAM6 }\citep{zhu_lgm_2022}\add{. CAM6-PaleoCalibr removed an inappropriate limiter on cloud ice number that is present in other versions of CAM6 and uses a shorter timestep in the MG2 microphysics scheme. CAM6-PaleoCalibr also uses a coarser resolution of 1.9$^\circ$x2.5$^\circ$ than the CAM6 PPE simulations, which have a resolution of 0.9$^\circ$x1.25$^\circ$. As a result of differences in horizontal resolution, a few of the default parameters are tuned differently, including a smaller \texttt{MG2\_DCS} value, a smaller dust emission factor, a smaller \texttt{CLUBB\_gamma}, and a larger seasalt emission scaling in the coarser version. In addition, the PaleoCalibr PPE uses year 1850 levels of atmospheric constituents and control climate SSTs whereas the CAM6 PPE uses year 2000 levels. Finally, the PaleoCalibr PPE uses a larger range of \texttt{CLUBB\_C8} and \texttt{CLUBB\_C11b} to explore their influence on the state dependence of the cloud feedback. The PaleoCalibr default simulation has a cloud feedback of 0.36 W m$^{-2}$ K$^{-2}$, which is smaller than that of the CAM6 PPE used above (0.56 W m$^{-2}$ K$^{-2}$). Figure S8 compares the assessed cloud feedbacks in the PaleoCalibr PPE to those in Figure }\ref{fig:assessed}\add{. Although the PaleoCalibr PPE default simulation has a smaller total cloud feedback than the CAM6 PPE default simulation, the relatively large high-cloud altitude feedback persists in the PaleoCalibr PPE, indicating that the changes made to reduce the total cloud feedback in the PaleoCalibr version do not address this discrepancy.}

\section{Discussion and conclusions} \label{sec:dc}

The CAM6 PPE provides a tool for evaluating the sensitivity of cloud feedbacks to uncertain model parameters. Surprisingly, the CAM6 PPE nearly captures the large spread across CMIP6 models.\add{ We did not expect this \emph{a priori}, although a similar spread was found in the Met Office's atmospheric GCM} \citep{rostron_impact_2020}. Further, the decomposition of these cloud feedbacks into 1) their SW and LW components, and 2) the cloud feedback components of \citet{sherwood_assessment_2020} show similar spreads across the two ensembles.\add{ Taking the spread across CMIP6 models to represent structural uncertainty and the spread across the CAM6 PPE to represent parametric uncertainty, these results indicate that the spread in cloud feedbacks due to parametric uncertainty is similar to that due to structural uncertainty, though this does not imply that changes in parameters explain differences across models. An advantage of using a PPE is that each parameter corresponds to a physical process, and the spread in cloud feedbacks due to each parameter and corresponding process can be estimated. The parameters with the greatest influence on cloud feedbacks across the CAM6 PPE are: \texttt{CLUBB\_C8}, \texttt{MG2\_DCS}, and \texttt{ZM\_capelmt}. These parameters control aspects of cloud thickness, cloud ice, and deep convection, respectively.}

We further use the CAM6 PPE to identify the parameters with the most influence on total cloud feedback and mean-state cloud errors. We find that the most influential parameters (by correlation) with mean-state cloud errors are MG2 accretion enhancement factor (\texttt{MG2\_accre\_enhan\_factor}) and ZM deep convective cloud fraction (\texttt{ZM\_cldfrac\_dp2}). We find that the most influential parameters (by correlation) with total cloud feedback are MG2 ice-snow autoconversion size threshold (\texttt{MG2\_DCS}), ZM triggering threshold for convection (\texttt{ZM\_capelmt}), and a CLUBB coefficient in the C8 skewness equation (\texttt{CLUBB\_C8}), and point out the values which are associated with the smallest cloud feedback errors.\add{ Therefore, improved representation of the processes corresponding to these parameters in models are promising avenues for constraining simulated cloud feedbacks in CAM6 and perhaps other models too. However,} \change{W}{w}e intentionally stop short of recommending these values for model \emph{tuning}. Firstly, climate model simulations of warmer climates should not be tuned to a desired `answer' since there are no observations with which to tune. Additionally, as mentioned above, the assessment of \citet{sherwood_assessment_2020}, while the most comprehensive to date, does not represent the truth. Therefore, climate models should not be developed or tuned to match these values. Further, the model may be subject to equifinality, that is, different combinations of parameters may yield the same cloud feedback error. Moreover, cloud feedbacks are not the only simulated feature of the model of interest but are the only subject of this analysis. Finally, it is clear from this analysis that ECS and cloud feedbacks are very sensitive to both parameter values are structural differences in the model, so do not provide a strong constraint for model tuning.

\remove{We evaluate the smaller cloud feedback in the CAM6 PPE default simulation as compared to CESM2 (AMIP). We find that changes to the model between CAM6.0 (used for CESM2 (AMIP)) and CAM6.3 (used for the CAM6 PPE ensemble) explain the reduced cloud feedback. It is unclear whether the decrease in cloud feedback between CAM6.0 and CAM6.3 foreshadows smaller cloud feedbacks in future versions of CESM because ultimately the model's cloud feedback will depend heavily on the role of coupling and further changes during model development.}

The range of high-cloud altitude feedback in the CAM6 PPE sits at larger values than that of CMIP6 models, AMIP models, and the WCRP assessment. More specifically, the CAM6 PPE simulations tend to have larger high-cloud altitude feedbacks than the assessed 1$\sigma$ range and than most CMIP6 and AMIP models, and CESM2 has the largest high-cloud altitude feedback of any CMIP6 model. In contrast, \citet{sherwood_assessment_2020} consider the high-cloud altitude feedback to be among the more certain of the six cloud feedback components. Their assessed value reflects the mean and standard deviation across GCMs, with support from observations, cloud resolving models, and large eddy simulations.\add{ In the CAM6 PPE, the parameters most correlated with the high-cloud altitude feedback are the convective parcel temperature perturbation parameter in the convection scheme (\texttt{ZM\_tiedke\_add}) and a scaling for subgrid velocity of ice activation in the microphysics scheme (\texttt{microp\_aero\_wsubi\_scale}).} \add{Further investigation of the relatively large high-cloud altitude feedback in CESM2 should be a priority for future investigations.}

\remove{To further understand our results, we compare the CAM6 PPE used here to another PPE which used the Paleoclimate-calibrated CAM6 (CAM6-PaleoCalibr), which has a smaller ECS than the standard version of CAM6 }\remove{. CAM6-PaleoCalibr removed an inappropriate limiter on cloud ice number that is present in the standard CAM6 and uses a shorter timestep in the MG2 microphysics scheme. CAM6-PaleoCalibr also uses a coarser resolution of 1.9$^\circ$x2.5$^\circ$ than that of standard CAM6, which has a resolution of 0.9$^\circ$x1.25$^\circ$. As a result of differences in horizontal resolution, a few of the default parameters are tuned differently, including a smaller MG2 DCS value, a smaller dust emission factor, a smaller CLUBB gamma, and a larger seasalt emission scaling in the coarser version. In addition, the PaleoCalibr PPE uses year 1850 levels of atmospheric constituents and control climate SSTs whereas the CAM6 PPE uses year 2000 levels. Finally, the PaleoCalibr PPE uses a larger range of CLUBB C8 and CLUBB C11b to explore their influence on the state dependence of the cloud feedback. The PaleoCalibr default simulation has a cloud feedback of 0.36 W m$^{-2}$ K$^{-2}$, which is smaller than that of the CAM6 PPE used above (0.56 W m$^{-2}$ K$^{-2}$). Figure S8 compares the assessed cloud feedbacks in the PaleoCalibr PPE to those in Figure }\remove{. Although the PaleoCalibr PPE default simulation has a smaller total cloud feedback than the CAM6 PPE default simulation, the relatively large high-cloud altitude feedback persists in the PaleoCalibr PPE, indicating that the changes made to reduce the total cloud feedback in the PaleoCalibr version do not address this discrepancy. Further investigation of the relatively large high-cloud altitude feedback in CESM2 should be a priority for future investigations.}

We use the CAM6 PPE to attempt to constrain the total cloud feedback by comparing 1) a measure of the cloud feedback discrepancy (as compared with \citet{sherwood_assessment_2020}) and 2) a measure of mean-state cloud errors in each CAM6 PPE ensemble member to the total cloud feedback. We find inconsistent results between the two metrics. Comparing with CMIP models, we find that WCRP assessed values provide a similar constraint in the CAM6 PPE as in the CMIP models (moderate cloud feedback). However, we find that despite the strong anticorrelation between mean-state cloud errors and total cloud feedback in CMIP models, mean-state cloud errors and total cloud feedback are uncorrelated in the CAM6 PPE\add{, which is broadly consistent with the results of }\citet{tsushima_investigating_2020}\add{, who used a PPE created with the Met Office's atmospheric GCM}. The mean-state cloud errors and difference from WCRP assessed values are uncorrelated with one another in both the CMIP and CAM6 PPE ensembles. While the comparison with WCRP assessed values provide similar constraints in both the CMIP and CAM6 PPE ensembles, this analysis is subject to several important limitations, both here and in \citet{zelinka_evaluating_2022}. \citet{sherwood_assessment_2020} estimate the uncertainty of each cloud feedback component, and this estimated uncertainty is not accounted for in the error metric used here. Further, the the assessment of \citet{sherwood_assessment_2020} has limitations, and more recent work has updated the assessed values (e.g. \citet{myers_observational_2021}). Finally, since the six cloud feedback components roughly sum to the total, the total cloud feedback estimate must approach that of \cite{sherwood_assessment_2020} as RMSD approaches zero, provided that the unassessed feedbacks are small. Therefore, we find our analysis which uses the CAM6 PPE to constrain total cloud feedback inconclusive.

%We find that simulations with the smallest differences have a total cloud feedback of about 0.5 to 0.7 W m$^{-2}$ K$^{-1}$. This is fairly consistent with the analysis of \citet{zelinka_evaluating_2022}, who did a similar analysis using CMIP5 and CMIP6 models and found total cloud feedbacks of about 0.4 to 0.6 W m$^{-2}$ K$^{-1}$ in the CMIP models with the smallest cloud feedback differences. However, this type of analysis has several limitations, both here and in \citet{zelinka_evaluating_2022}. \citet{sherwood_assessment_2020} estimate the uncertainty of each cloud feedback component, and this estimated uncertainty is not accounted for in the error metric used here. Further, the the assessment of \citet{sherwood_assessment_2020} has limitations, and more recent work has updated the assessed values (e.g. \citet{myers_observational_2021}). Finally, we find that the range of cloud feedbacks estimated by the CAM6 PPE corresponds to an ECS of approximately 3 to 4 K, but this is likely an underestimate due to the lack of pattern effect in the CAM6 PPE set up. Our estimate of ECS is lower than that of \citet{peatier_investigating_2022}, who used a CNRM-CM6-1 PPE with a \textit{patterned warming}, and thus the ability to capture a pattern effect, to estimate an ECS range of 4.26 to 6.15 K, using different methods.

Given the substantial influence of parameters on cloud feedbacks in the CAM6 PPE, we ask whether parameter tuning is responsible for the increase in cloud feedback and thus ECS from CAM5 to CAM6. Using a simple regression model, we estimate that changes in parameters between CAM5 and CAM6 have little influence on the total cloud feedback, and therefore are not responsible for the increase in cloud feedback between generations. This indicates that structural changes to the model between CAM5 and CAM6 are responsible for the increase in total cloud feedback and ECS.

\add{We evaluate the smaller cloud feedback in the CAM6 PPE default simulation as compared to CESM2 (AMIP). We find that changes to the model between CAM6.0 (used for CESM2 (AMIP)) and CAM6.3 (used for the CAM6 PPE ensemble) explain the reduced cloud feedback. It is unclear whether the decrease in cloud feedback between CAM6.0 and CAM6.3 foreshadows smaller cloud feedbacks in future versions of CESM because ultimately the model's cloud feedback will depend heavily on the role of coupling and further changes during model development.}

Overall, our analysis highlights the large sensitivity of ECS estimates to both changes in parameters and structural changes in CAM6, which may help explain why it has been so difficult to constrain.

%%%%%%%%%%%%%%%%%%%%%%%%%%%%%%%%%%%%%%%%%%%%%%%%%%%%%%%%%%%%%%%%%%%%%
% ACKNOWLEDGMENTS
%%%%%%%%%%%%%%%%%%%%%%%%%%%%%%%%%%%%%%%%%%%%%%%%%%%%%%%%%%%%%%%%%%%%%
\acknowledgments
This work was funded by NOAA MAPP under award NA20OAR4310392. BM acknowledges support by the U.S. Department of Energy under Award Number DE-SC0022070 and National Science Foundation (NSF) IA 1947282. This material is based upon work supported by the National Center for Atmospheric Research, which is a major facility sponsored by the National Science Foundation under Cooperative Agreement No. 1852977. Computing resources were provided by the Climate Simulation Laboratory at NCAR’s Computational and Information Systems Laboratory (CISL). The authors thank Jiang Zhu for helpful conversations and access to the PaleoCalibr PPE data, Jesse Nusbaumer for help with running CAM6.0, \add{and three anonymous reviewers for their helpful feedback.}

%%%%%%%%%%%%%%%%%%%%%%%%%%%%%%%%%%%%%%%%%%%%%%%%%%%%%%%%%%%%%%%%%%%%%
% DATA AVAILABILITY STATEMENT
%%%%%%%%%%%%%%%%%%%%%%%%%%%%%%%%%%%%%%%%%%%%%%%%%%%%%%%%%%%%%%%%%%%%%
% 
%
\datastatement

The \add{CAM6 }PPE dataset is available at https://doi.org/10.26024/bzne-yf09 \citep{eidhammer_cesm22-cam6_2022}. The CMIP and AMIP data are from Mark Zelinka's github repositories, as described in Section \ref{sec:data} and are available at https://github.com/mzelinka/cmip56\_forcing\_feedback\_ecs and https://github.com/mzelinka/assessed-cloud-fbks.

%%%%%%%%%%%%%%%%%%%%%%%%%%%%%%%%%%%%%%%%%%%%%%%%%%%%%%%%%%%%%%%%%%%%%
% REFERENCES
%%%%%%%%%%%%%%%%%%%%%%%%%%%%%%%%%%%%%%%%%%%%%%%%%%%%%%%%%%%%%%%%%%%%%
 %This shows how to enter the commands for making a bibliography using
 %BibTeX. It uses references.bib and the ametsoc2014.bst file for the style.

\bibliographystyle{ametsocV6}
\bibliography{PPE}

\end{document}